\documentclass[
    twocolumn,
    times
    ]{aastex63}
\usepackage[utf8]{inputenc}
\usepackage[T1]{fontenc}

\usepackage{upgreek}
\usepackage{commath}
\usepackage{rotating}
\usepackage[super]{nth}
\DeclareMathOperator*{\argmax}{arg\,max}

\graphicspath{{figures/}}

\let\tablenum\relax
\usepackage[binary-units]{siunitx}[=v2]
\DeclareSIUnit{\arcmin}{arcmin}
\DeclareSIUnit{\parsec}{pc}
\DeclareSIUnit{\arcsec}{arcsec}
\DeclareSIUnit{\mas}{mas}
\DeclareSIUnit{\jansky}{Jy}
\DeclareSIUnit{\FRBs}{FRBs}
\DeclareSIUnit{\FRB}{FRB}
\DeclareSIUnit{\sky}{sky}
\DeclareSIUnit{\day}{day}
\DeclareSIUnit{\month}{month}
\DeclareSIUnit{\ethernet}{E}
\DeclareSIUnit{\deg}{deg}
\DeclareSIUnit{\TECu}{TECu}
\DeclareSIUnit{\electron}{electron}
\DeclareSIUnit{\MSPS}{MSPS}

\shorttitle{VLBI with the ARO 10-m Radio Telescope}
\shortauthors{Cassanelli et al}

\begin{document}
\title{Localizing FRBs through VLBI with the Algonquin Radio Observatory 10-m Telescope}


\author[0000-0003-2047-5276]{T.~Cassanelli}
  \affiliation{David A.~Dunlap Department of Astronomy \& Astrophysics, University of Toronto, 50 St.~George Street, Toronto, ON M5S 3H4, Canada}
  \affiliation{Dunlap Institute for Astronomy \& Astrophysics, University of Toronto, 50 St.~George Street, Toronto, ON M5S 3H4, Canada}

\author[0000-0002-4209-7408]{Calvin Leung}
  \affiliation{MIT Kavli Institute for Astrophysics and Space Research, Massachusetts Institute of Technology, 77 Massachusetts Ave, Cambridge, MA 02139, USA}
  \affiliation{Department of Physics, Massachusetts Institute of Technology, 77 Massachusetts Ave, Cambridge, MA 02139, USA}

\author[0000-0003-1842-6096]{M.~Rahman}
  \affiliation{Sidrat Research, PO Box 73527 RPO Wychwood, Toronto, ON M6C 4A7, Canada}

\author[0000-0003-4535-9378]{K.~Vanderlinde}
  \affiliation{David A.~Dunlap Department of Astronomy \& Astrophysics, University of Toronto, 50 St.~George Street, Toronto, ON M5S 3H4, Canada}
  \affiliation{Dunlap Institute for Astronomy \& Astrophysics, University of Toronto, 50 St.~George Street, Toronto, ON M5S 3H4, Canada}

\author[0000-0002-0772-9326]{J.~Mena-Parra}
  \affiliation{MIT Kavli Institute for Astrophysics and Space Research, Massachusetts Institute of Technology, 77 Massachusetts Ave, Cambridge, MA 02139, USA}

\author[0000-0003-1860-1632]{S.~Cary}
  \affiliation{MIT Kavli Institute for Astrophysics and Space Research, Massachusetts Institute of Technology, 77 Massachusetts Ave, Cambridge, MA 02139, USA}
  \affiliation{Department of Astronomy, Wellesley College, 106 Central Street, Wellesley, MA 02481, USA}

\author[0000-0002-4279-6946]{Kiyoshi W.~Masui}
  \affiliation{MIT Kavli Institute for Astrophysics and Space Research, Massachusetts Institute of Technology, 77 Massachusetts Ave, Cambridge, MA 02139, USA}
  \affiliation{Department of Physics, Massachusetts Institute of Technology, 77 Massachusetts Ave, Cambridge, MA 02139, USA}

\author[0000-0001-5373-5914]{Jing Luo}
  \affiliation{Canadian Institute for Theoretical Astrophysics, 60 St.~George Street, Toronto, ON M5S 3H8, Canada}

\author[0000-0001-7453-4273]{H.-H.~Lin}
  \affiliation{Institute of Astronomy and Astrophysics, Academia Sinica, Astronomy-Mathematics Building, No.~1, Sec.~4, Roosevelt Road, Taipei 10617, Taiwan}
  \affiliation{Canadian Institute for Theoretical Astrophysics, 60 St.~George Street, Toronto, ON M5S 3H8, Canada}

\author[0000-0001-7505-5223]{A.~Bij}
  \affiliation{Canadian Institute for Theoretical Astrophysics, 60 St.~George Street, Toronto, ON M5S 3H8, Canada}
  \affiliation{Department for Physics, Engineering Physics and Astrophysics, Queen’s University, Kingston, ON, K7L 3N6, Canada}

\author[0000-0002-3937-4662]{A.~Gill}
  \affiliation{David A.~Dunlap Department of Astronomy \& Astrophysics, University of Toronto, 50 St.~George Street, Toronto, ON M5S 3H4, Canada}
  \affiliation{Dunlap Institute for Astronomy \& Astrophysics, University of Toronto, 50 St.~George Street, Toronto, ON M5S 3H4, Canada}

\author[0000-0001-7888-3470]{D.~Baker}
  \affiliation{Canadian Institute for Theoretical Astrophysics, 60 St.~George Street, Toronto, ON M5S 3H8, Canada}


\author[0000-0003-3772-2798]{Kevin Bandura}
  \affiliation{Lane Department of Computer Science and Electrical Engineering, 1220 Evansdale Drive, PO Box 6109 Morgantown, WV 26506, USA}
  \affiliation{Center for Gravitational Waves and Cosmology, West Virginia University, Chestnut Ridge Research Building, Morgantown, WV 26505, USA}
\author[0000-0002-4064-7883]{S.~Berger}
  \affiliation{Department of Physics, McGill University, 3600 rue University, Montr\'eal, QC H3A 2T8, Canada}
  \affiliation{McGill Space Institute, McGill University, 3550 rue University, Montr\'eal, QC H3A 2A7, Canada}
\author[0000-0001-8537-9299]{P.~J.~Boyle}
  \affiliation{Department of Physics, McGill University, 3600 rue University, Montr\'eal, QC H3A 2T8, Canada}
  \affiliation{McGill Space Institute, McGill University, 3550 rue University, Montr\'eal, QC H3A 2A7, Canada}
\author[0000-0002-1800-8233]{Charanjot Brar}
  \affiliation{Department of Physics, McGill University, 3600 rue University, Montr\'eal, QC H3A 2T8, Canada}
  \affiliation{McGill Space Institute, McGill University, 3550 rue University, Montr\'eal, QC H3A 2A7, Canada}
\author[0000-0002-2878-1502]{S.~Chatterjee}
  \affiliation{Cornell Center for Astrophysics and Planetary Science, Ithaca, NY 14853, USA}
\author[0000-0003-2319-9676]{D.~Cubranic}
  \affiliation{Department of Physics and Astronomy, University of British Columbia, 6224 Agricultural Road, Vancouver, BC V6T 1Z1 Canada}
\author[0000-0001-7166-6422]{Matt Dobbs}
  \affiliation{Department of Physics, McGill University, 3600 rue University, Montr\'eal, QC H3A 2T8, Canada}
  \affiliation{McGill Space Institute, McGill University, 3550 rue University, Montr\'eal, QC H3A 2A7, Canada}
\author[0000-0001-8384-5049]{E.~Fonseca}
  \affiliation{Department of Physics, McGill University, 3600 rue University, Montr\'eal, QC H3A 2T8, Canada}
  \affiliation{McGill Space Institute, McGill University, 3550 rue University, Montr\'eal, QC H3A 2A7, Canada}
  \affiliation{Department of Physics and Astronomy, West Virginia University, P.O. Box 6315, Morgantown, WV 26506, USA }
  \affiliation{Center for Gravitational Waves and Cosmology, West Virginia University, Chestnut Ridge Research Building, Morgantown, WV 26505, USA}
\author[0000-0003-1884-348X]{D.~C.~Good}
  \affiliation{Department of Physics and Astronomy, University of British Columbia, 6224 Agricultural Road, Vancouver, BC V6T 1Z1 Canada}
\author[0000-0003-4810-7803]{J.~F.~Kaczmarek}
  \affiliation{Dominion Radio Astrophysical Observatory, Herzberg Research Centre for Astronomy and Astrophysics, National Research Council Canada, PO Box 248, Penticton, BC V2A 6J9, Canada}
\author[0000-0001-9345-0307]{V.~M.~Kaspi}
  \affiliation{Department of Physics, McGill University, 3600 rue University, Montr\'eal, QC H3A 2T8, Canada}
  \affiliation{McGill Space Institute, McGill University, 3550 rue University, Montr\'eal, QC H3A 2A7, Canada}
\author[0000-0003-1455-2546]{T.~L.~Landecker}
  \affiliation{Dominion Radio Astrophysical Observatory, Herzberg Research Centre for Astronomy and Astrophysics, National Research Council Canada, PO Box 248, Penticton, BC V2A 6J9, Canada}
\author[0000-0003-2116-3573]{A.~E.~Lanman}
  \affiliation{Department of Physics, McGill University, 3600 rue University, Montr\'eal, QC H3A 2T8, Canada}
\author[0000-0001-7931-0607]{Dongzi Li}
  \affiliation{Cahill Center for Astronomy and Astrophysics, California Institute of Technology, 1216 E California Boulevard, Pasadena, CA 91125, USA}
\author[0000-0002-2885-8485]{J.~W.~McKee}
  \affiliation{Canadian Institute for Theoretical Astrophysics, 60 St.~George Street, Toronto, ON M5S 3H8, Canada}
\author[0000-0001-8845-1225]{B.~W.~Meyers}
  \affiliation{Department of Physics and Astronomy, University of British Columbia, 6224 Agricultural Road, Vancouver, BC V6T 1Z1 Canada}
\author[0000-0002-2551-7554]{D.~Michilli}
  \affiliation{Department of Physics, McGill University, 3600 rue University, Montr\'eal, QC H3A 2T8, Canada}
  \affiliation{McGill Space Institute, McGill University, 3550 rue University, Montr\'eal, QC H3A 2A7, Canada}
\author[0000-0002-9225-9428]{Arun Naidu}
  \affiliation{Department of Physics, McGill University, 3600 rue University, Montr\'eal, QC H3A 2T8, Canada}
  \affiliation{McGill Space Institute, McGill University, 3550 rue University, Montr\'eal, QC H3A 2A7, Canada}
\author[0000-0002-3616-5160]{Cherry Ng}
  \affiliation{Dunlap Institute for Astronomy \& Astrophysics, University of Toronto, 50 St.~George Street, Toronto, ON M5S 3H4, Canada}
\author[0000-0003-3367-1073]{Chitrang Patel}
  \affiliation{Department of Physics, McGill University, 3600 rue University, Montr\'eal, QC H3A 2T8, Canada}
  \affiliation{Dunlap Institute for Astronomy \& Astrophysics, University of Toronto, 50 St.~George Street, Toronto, ON M5S 3H4, Canada}
\author[0000-0002-8912-0732]{Aaron~B.~Pearlman}
  \altaffiliation{McGill Space Institute~(MSI) Fellow.}
  \affiliation{Department of Physics, McGill University, 3600 rue University, Montr\'eal, QC H3A 2T8, Canada }
  \altaffiliation{FRQNT Postdoctoral Fellow.}
  \affiliation{McGill Space Institute, McGill University, 3550 rue University, Montr\'eal, QC H3A 2A7, Canada }
\author[0000-0003-2155-9578]{U.~L.~Pen}
  \affiliation{Institute of Astronomy and Astrophysics, Academia Sinica, Astronomy-Mathematics Building, No.~1, Sec.~4, Roosevelt Road, Taipei 10617, Taiwan}
  \affiliation{Canadian Institute for Theoretical Astrophysics, 60 St.~George Street, Toronto, ON M5S 3H8, Canada}
  \affiliation{Canadian Institute for Advanced Research, 661 University Ave, Toronto, ON M5G 1M1}
  \affiliation{Dunlap Institute for Astronomy \& Astrophysics, University of Toronto, 50 St.~George Street, Toronto, ON M5S 3H4, Canada}
  \affiliation{Perimeter Institute for Theoretical Physics, 31 Caroline Street N, Waterloo, ON N25 2YL, Canada}
\author[0000-0002-4795-697X]{Ziggy Pleunis}
  \affiliation{Department of Physics, McGill University, 3600 rue University, Montr\'eal, QC H3A 2T8, Canada}
  \affiliation{McGill Space Institute, McGill University, 3550 rue University, Montr\'eal, QC H3A 2A7, Canada}
\author[0000-0002-7326-2779]{Brendan Quine}
  \affiliation{Thoth Technology Inc.~33387 Highway 17, Deep River, Ontario K0J 1P0, Canada}
  \affiliation{Department of Physics and Astronomy, York University, 4700 Keele St, Toronto, Ontario M3J 1P3, Canada}
\author[0000-0003-3463-7918]{A.~Renard}
  \affiliation{Dunlap Institute for Astronomy \& Astrophysics, University of Toronto, 50 St.~George Street, Toronto, ON M5S 3H4, Canada}
\author[0000-0001-5504-229X]{Pranav Sanghavi}
  \affiliation{Lane Department of Computer Science and Electrical Engineering, 1220 Evansdale Drive, PO Box 6109 Morgantown, WV 26506, USA}
  \affiliation{Center for Gravitational Waves and Cosmology, West Virginia University, Chestnut Ridge Research Building, Morgantown, WV 26505, USA}
\author[0000-0002-2088-3125]{K.~M.~Smith}
  \affiliation{Perimeter Institute for Theoretical Physics, 31 Caroline Street N, Waterloo, ON N25 2YL, Canada}
\author[0000-0001-9784-8670]{Ingrid Stairs}
  \affiliation{Department of Physics and Astronomy, University of British Columbia, 6224 Agricultural Road, Vancouver, BC V6T 1Z1 Canada}
\author[0000-0003-2548-2926]{Shriharsh P.~Tendulkar}
  \affiliation{Department of Astronomy and Astrophysics, Tata Institute of Fundamental Research, Mumbai, 400005, India}
  \affiliation{National Centre for Radio Astrophysics, Post Bag 3, Ganeshkhind, Pune, 411007, India}

\begin{abstract}
    The CHIME/FRB experiment has detected thousands of Fast Radio Bursts (FRBs) due to its sensitivity and wide field of view; however, its low angular resolution prevents it from localizing events to their host galaxies.
    Very Long Baseline Interferometry (VLBI), triggered by FRB detections from CHIME/FRB will solve the challenge of localization for non-repeating events.
    Using a refurbished 10-m radio dish at the Algonquin Radio Observatory located in Ontario Canada, we developed a testbed for a VLBI experiment with a theoretical $\lambda/D\lesssim\SI{30}{\mas}$.
    We provide an overview of the 10-m system and describe its refurbishment, the data acquisition, and a procedure for  fringe fitting that simultaneously estimates the geometric delay used for localization and the dispersive delay from the ionosphere. Using single pulses from the Crab pulsar, we validate the system and localization procedure, and analyze the clock stability between sites, which is critical for coherently delay-referencing an FRB event.
    We find a localization of \SI{\sim200}{\mas} is possible with the performance of the current system (single-baseline). Furthermore, for sources with insufficient signal or restricted wideband to simultaneously measure both geometric and ionospheric delays, we show that the differential ionospheric contribution between the two sites must be measured to a precision of \SI{1e-8}{\parsec\per\centi\m\cubed} to provide a reasonable localization from a detection in the \SIrange{400}{800}{\mega\hertz} band.
    Finally we show detection of an FRB observed simultaneously in the CHIME and the Algonquin 10-m telescope, the first non-repeating FRB in this long baseline.
    This project serves as a testbed for the forthcoming CHIME/FRB Outriggers project.
\end{abstract}

\correspondingauthor{Tomas Cassanelli}
\email{cassanelli@astro.utoronto.ca}


\section{Introduction}
\label{sec:intro}

Fast Radio Bursts are bright, millisecond-long flashes of radio emission that were first discovered by \cite{2007Sci...318..777L} and have been observed to be distributed throughout the sky \citep{2021arXiv210604353J}. Their origins are among the key unresolved questions in astrophysics, despite hundreds of FRBs having been observed to date \citep{2016PASA...33...45P,2021arXiv210604352T}.
A significant challenge towards understanding FRBs lies in our inability to determine their precise on-sky positions, which is critical for identifying their host galaxies \citep{2017ApJ...849..162E}. Despite the fact that some facilities are able to localize FRBs in limited numbers, localization remains a challenging problem (with the exception of low redshift nearby events \citeauthor{2021ApJ...910L..18B} \citeyear{2021ApJ...910L..18B}). This is mainly due to the fact that most FRBs are non-repeating, and a large field of view is required in order to increase the probability of detection.
A few astronomical facilities are able to identify single-burst FRBs' host galaxies \citep{2017ApJ...841L..12B, 2019Sci...365..565B, 2019Natur.572..352R}, but their detection rates and fields of view are low.
By achieving more precise localizations of non-repeating FRBs, we can constrain FRB populations \citep{2017ApJ...834L...8M, 2021arXiv210511445K} and improve the classification for repeaters and non-repeaters, and also put them to use as probes for fundamental astrophysics \citep{2019A&ARv..27....4P}.

The Canadian Hydrogen Intensity Mapping Experiment (CHIME) is a radio telescope operating in the \SIrange{400}{800}{\mega\hertz} band, with an \SI{8000}{\meter\squared} aperture area of semi-cylindrical paraboloid reflectors, located at the Dominion Radio Astrophysical Observatory (DRAO) near Penticton, British Columbia.
Each of the four cylinders is instrumented with \num{256} cloverleaf feeds suspended along the cylindrical axis, and observers a \SI{>200}{\deg\squared} Field of View (FoV). The combination of the FoV, wide bandwidth, large aperture area, and a powerful correlator makes CHIME a great tool for FRB detection. The CHIME/FRB Collaboration \citep{2018ApJ...863...48C} has already published the detection of \num{21} repeating and \num{474} non-repeating FRBs \citep{2021arXiv210604352T}.
CHIME/FRB and its baseband system are able to localize FRBs with a resolution higher than $\frac{\lambda}{D}$\footnote{Compared to a standard circular aperture with diameter $D$ observing at a wavelength $\lambda$.}, and as shown in \citep{2019ApJ...879...16M}, with a best case limit of \SI{\sim1}{\arcmin} \citep{2020arXiv201006748M}. In order to determine the host galaxies of FRBs, however, localizations better than \SI{1}{\arcsec} are required.

VLBI is a technique that combines observations from multiple separated telescopes, effectively turning them into one single telescope with an enhanced angular resolution that scales inversely with the projected baseline length. This technique is one method by which we can improve the diffraction-limited beamwidth of CHIME/FRB's localizations: by correlating baseband data between CHIME and another site, we can use VLBI to find fringes for a single FRB event and coherently delay-reference to a nearby calibrator, hence improving the CHIME/FRB localizations by \num{2} to \num{3} orders of magnitude.
So far only repeaters have been localized with VLBI scheduled observations \citep{2020Natur.577..190M} and not one-off events.

\begin{figure}[t]
    \centering
    \includegraphics{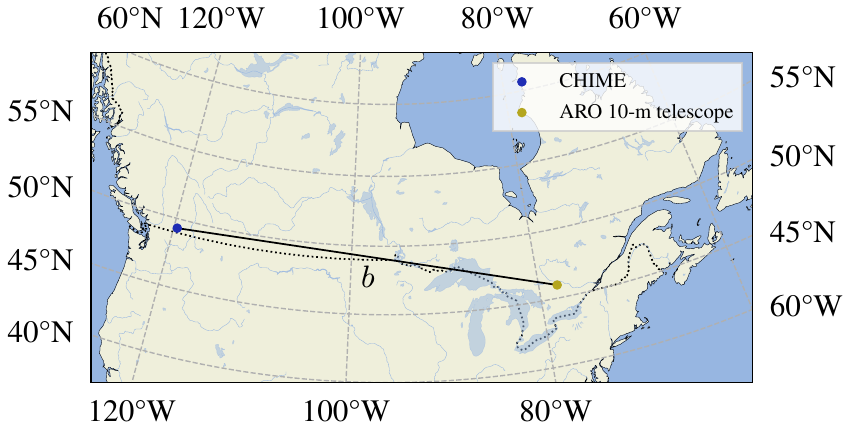}
    \caption{VLBI projection of the DRAO and ARO baseline. The black straight line represents the baseline $b$ from Penticton BC to Algonquin ON, with a distance of over \SI{3000}{\kilo\m}.}
    \label{fig:vlbi_projection}
\end{figure}

In this paper we describe VLBI between CHIME and a 10-m telescope at the Algonquin Radio Observatory (ARO). ARO is \SI{260}{\kilo\m} West of Ottawa in Algonquin Park, Ontario Canada. It was established in 1963 by the National Research Council Canada (NRC), and its major instrument was a 46-m radio telescope. The 10-m telescope, a prime-focus paraboloid, was constructed at the same time. The baseline between CHIME and ARO is over \SI{3000}{\kilo\m}, offering a diffraction limited resolution within $\theta\lesssim\SI{30}{\mas}$. The Earth projection of the baseline is shown in Figure \ref{fig:vlbi_projection}.

The CHIME/FRB Collaboration is actively developing a new set of CHIME Outrigger, composed of cylindrical telescopes at distances of one \num{\sim20} to several thousand kilometers from the CHIME telescope. The ARO 10-m telescope dish and the CHIME Pathfinder \citep{2021AJ....161...81L} serve as testbeds for the CHIME/FRB Outriggers project. Along with CHIME, the outriggers will perform an autonomous VLBI survey to localize over a thousand FRBs with \SI{50}{\mas} precision. Outriggers will cover mostly the same area that CHIME does (with up to three baselines), and correlate beams digitally while preserving coherence over the FRB sweep. Further, near in-beam calibration can be performed while digitally pointing to different locations on sky, with only an angular space interpolation (rather than over time; clock system), and eventually will be expanded to a grid of pulsar calibrators.
The outriggers project's aim is to localize one-off FRBs with a VLBI network particularly designed for this purpose, different from a classical VLBI imaging facility, where the pulse is only visible over $\sim$\si{\milli\s} time scales.
The ARO 10-m testbed is the initial effort to correlate FRBs for the upcoming project.

This paper will proceed as follows. We provide an overview of ARO, the 10-m telescope, and our new instrumentation of the telescope in Section \ref{sec:instrumentation}. In Section \ref{sec:vlbi_between_the_aro_10-m_and_chime} we describe VLBI experiments between CHIME and the ARO 10-m telescope. Section \ref{sec:early_science_results} describes early science results, including the detection of VLBI fringes from one FRB. We discuss our results in Section \ref{sec:discussion} and present our conclusions in Section \ref{sec:conc}.
Appendix \ref{sec:rf_chain} details the 10-m radio frequency chain, and Appendix \ref{sec:visibility_and_cross-correlation_in_baseband_space} provides a detailed explanation of visibilities and cross-correlation with baseband data.

\section{Instrumentation}
\label{sec:instrumentation}

\subsection{The Algonquin Radio Observatory}
\label{sec:aro}

We note that the very first VLBI observations were made between DRAO and ARO in \citeyear{1967Sci...156.1592B} by \citeauthor{1967Sci...156.1592B} ARO provides three benefits for VLBI, a site thousands of kilometers from CHIME at DRAO, a clean Radio Frequency Interference (RFI) environment, and access to a hydrogen maser time standard. Although built by NRC, the ARO 46-m Telescope is currently operated by a private company, THOTH\footnote{For more information visit: \url{http://thothx.com}.}. The 10-m telescope was not in use at the time we began our work, and had not been used for many years. To provide a testbed for development of VLBI techniques for the CHIME/FRB Outrigger project, we equipped the 10-m telescope with a new feed and receiver system, sending telescope signals via \SI{200}{\m} of buried coaxial cables (from the 1960s) to our new digital processing system in the control room of the 46-m telescope. The \SI{10}{\m} aperture provides sufficient gain to be a useful complement to CHIME.

\subsection{The 10-m telescope}
\label{sec:the_10-m_telescope}

\begin{figure}[t]
    \centering
    \includegraphics[width=.4\textwidth]{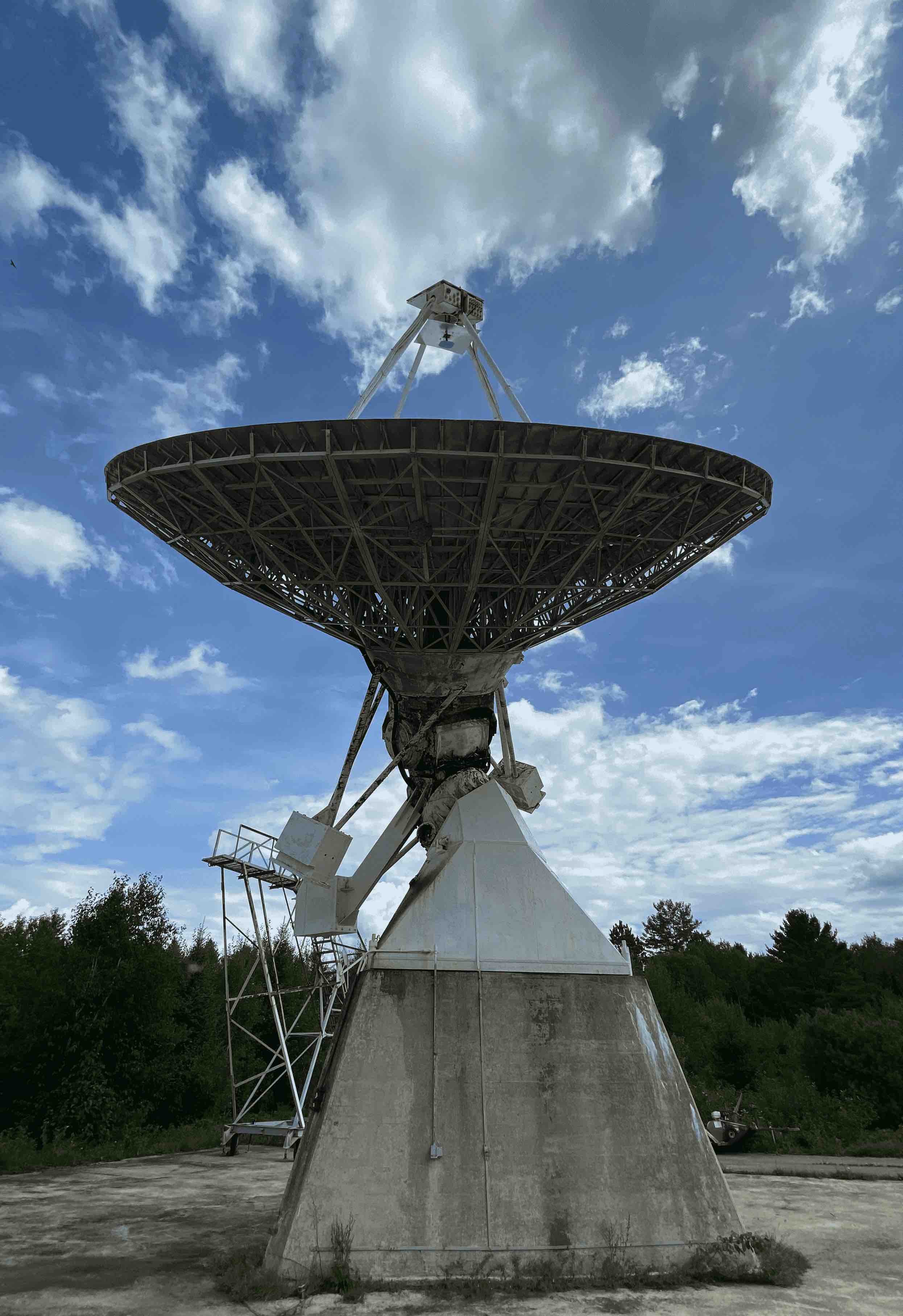}
    \caption{ARO 10-m telescope; photograph taken in September 2020. The backplane and CHIME feed are visible at the telescope's focus. The feed in the paraboloid focus is slightly offset towards the vertex.}
    \label{fig:10m}
\end{figure}

The 10-m telescope (Figure \ref{fig:10m}) was inaugurated in 1964 \citep{1961P&SS....5..307M, National-Research-Council-of-Canada:1969aa}, with an equatorial mount and a manual mechanical engine with declination and hour angle drive (or polar drive). 
The paraboloidal dish has a structure with four feed-support struts to support the focus equipment, with a surface accuracy of \SI{0.063}{\centi\meter} rms at the time of its construction \citep{Dawson:1970aa}. Coaxial cables with standard BNC and N-type lines run through the feed-support struts and connect the feed and first stage amplifiers (power and analog cables).
The telescope is equipped with an analog system for pointing and a polar axis drive unit located physically at the telescope exterior.

Table \ref{tab:ARO10m_prop} gives specifications of the telescope and its location. The declination and polar drive and their indicator systems were only partly functional at the beginning of our work, but we were able to point the telescope by applying external torque to the gearboxes. The telescope was set to the declination of Taurus A to receive pulses from PSR B05321+21, which we used as a test source. The hour angle was set West of the ARO meridian so that area of sky seen by the main beam coincided with that seen by CHIME on the meridian at DRAO (see Figure \ref{fig:fov}). The azimuth and elevation of this position are \SI{248}{\deg} and \SI{50}{\deg} respectively. We estimate the accuracy of this pointing to be \SI{\sim1}{\deg}, checked by the PSR B05321+21 transit through the telescope beam.

\begin{table}[t]
    \centering
    \caption{ARO 10-m telescope properties}
    \label{tab:ARO10m_prop}
    \begin{tabular}{ll}
        \tableline
        Telescope mount & Equatorial RA and DEC \\
        Dish diameter & \SI{10}{\meter} \\
        $f/D$ & \num{0.4} \\
        Surface accuracy & \SI{0.063}{\centi\meter} rms\\
        Latitude & \SI{45.955}{\deg} \\
        Longitude & \SI{-78.070}{\deg} \\
        Altitude & $\SI{\sim50}{\deg}$ \\
        Azimuth & $\SI{\sim248}{\deg}$ \\
        \tableline
    \end{tabular}
\end{table}

We installed a new analog system on the 10-m telescope, including a new feed and low-noise amplifiers (LNAs) at the focus, and second-stage amplification in the telescope pedestal. We used underground coaxial cables, installed in the 1960s, to connect to a new digital acquisition system in the control room of the 46-m telescope, \SI{200}{\m} away. See Section \ref{sec:analog_signal_chain} and Appendix \ref{sec:rf_chain} for details of the analog system, and Section \ref{sec:digital_processing} for a description of the digital system.

With the completion of these upgrades, the telescope is fully operational and working 24/7 with occasional power interruptions during the winter months due to the site's remote location.
Its day-to-day operations and monitoring are done remotely over a satellite-based Internet connection.

\subsection{Analog Signal Chain}
\label{sec:analog_signal_chain}

The decommissioned 10-m telescope, along with its analog chain and cabling, had fallen into disuse prior to beginning this project. Before this project commenced, the system cables were tested and confirmed to still be in good working order but with attenuation due to the age of the cables; the existing connectors were also improved and signal amplification was added. Additionally, the entire backend and partial frontend systems were updated.
The receiver was equipped with a dual linear polarized feed, identical to those used to equip CHIME \citep{2017arXiv170808521D}. The receiver has a working bandwidth of \SI{400}{\mega\hertz} and has a noise injection SMA connection, used for on site tests and telescope characterization. The feed was placed in the telescope's focus and connected to the first stage amplifiers (which are the same as those on CHIME).

\subsubsection{Radio frequency signal chain}
\label{sec:radio_frequency_signal_chain}
There are three amplifier stages: $\mathcal{S}_1$, $\mathcal{S}_2$ and $\mathcal{S}_3$. The first stage amplifiers are located right next to the cloverleaf feed (telescope's focus) with an SMA connection. The first stage amplifiers are connected by dual polarization lines to the second stage amplifiers in the telescope's pedestal. The pedestal also contains the power supplies for $\mathcal{S}_1$ and $\mathcal{S}_2$. The final stage, $\mathcal{S}_3$, is at the end of the underground connected lines from the telescope's pedestal to the control room. In the control room there is another set of amplifiers right before the digitizer ICE system \citep{2016JAI.....541005B}.

\begin{figure}[t]
    \centering
    \includegraphics{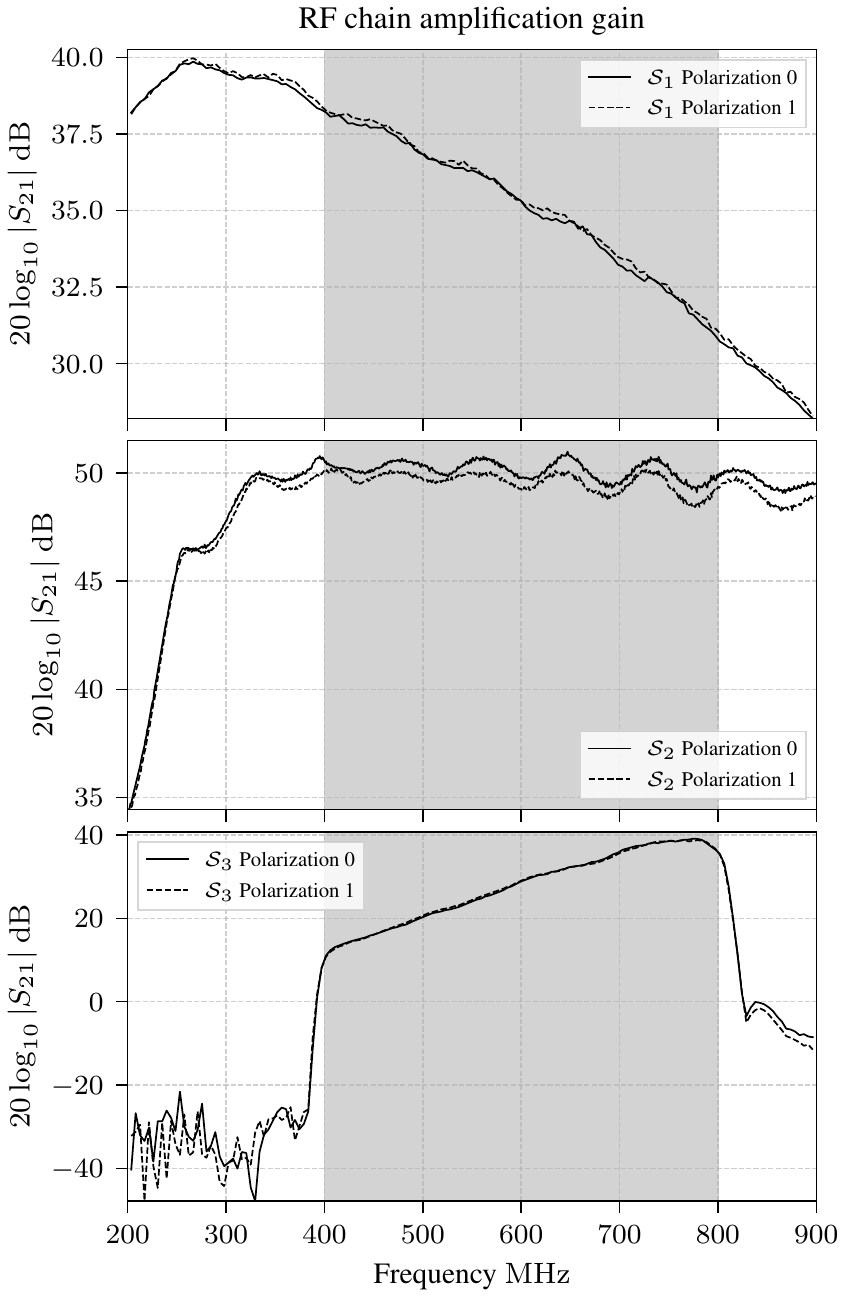}
    \caption{Gains of the amplifiers in stages $\mathcal{S}_1$, $\mathcal{S}_2$, and $\mathcal{S}_3$. Each of the amplifiers was measured and tested on site with a Vector Network Analyzer. The grey area is the bandwidth \SIrange{400}{800}{\mega\hertz}. The $y$-axis gain $G$ is presented in terms of the scattering parameter or S-parameter, $\envert{G}=\envert{{S}_{21}}$.
    Data were taken during the feed installation in April 2019. The strong ripple in the amplifier response seen in the middle panel is due to cable reflections.}
    \label{fig:s_amps}
\end{figure}

The Low Noise Amplifiers (LNAs) in $\mathcal{S}_1$ are custom built CHIME amplifiers (see the top graph in Figure \ref{fig:s_amps}). Each amplifier is bounded with a Radio Frequency (RF) shield enclosure that also protects for weather conditions.

Due to the large distance spanned by the connection between the 10-m dish and the control room, an extra amplification was required from $\mathcal{S}_2$ to $\mathcal{S}_3$. These are commercial LNAs placed consecutively and fed with the analog lines that come from the telescope focus in $\mathcal{S}_1$ (see the middle panel in Figure \ref{fig:s_amps}).

$\mathcal{S}_3$ contains the last set of amplifiers after the long underground lines (\SI{\sim200}{\m}) and before the signal digitalization. This last set of amplifiers is composed of custom CHIME Pathfinder amplifiers \citep{10.1117/12.2054950}, a set of commercial line equalizers, and \SIrange{400}{800}{\mega\hertz} bandpass filters. The $\mathcal{S}_3$ amplifiers' gain (bottom panel of Figure \ref{fig:s_amps}) compensate for the gain drop in $\mathcal{S}_1$.

For a detailed description of components and the analog chain see Appendix \ref{sec:rf_chain}.

\subsection{Digital Processing}
\label{sec:digital_processing}

The digital processing at the ARO 10-m telescope is similar to that done by CHIME/FRB.
The analog signal arrives at the ADCs on the ICE, which are clocked by a hydrogen maser, and an initial timestamp for each acquisition is provided by a Global Positioning System (GPS) signal. The underground analog signals (two polarizations) come from the 10-m, and two extra SMA connections on the board are related to the maser and GPS timestamps.
Following this, the same main configuration from CHIME/FRB is applied, i.e., we sample raw voltages at \SI{8}{\bit}, apply a Polyphase Filter Bank (PFB; with an identical number of taps) \citep{2016arXiv160703579P} at \SI{18}{\bit}, and then quantize to \SI[parse-numbers=false]{4+4}{\bit} resolution. After channelization is done baseband data has \num{1024} frequency channels spanning from \SIrange{400}{800}{\mega\hertz} and a sample rate of \SI{390.625}{\kilo\hertz}.

\begin{figure}[t]
    \centering
    \includegraphics[width=.47\textwidth]{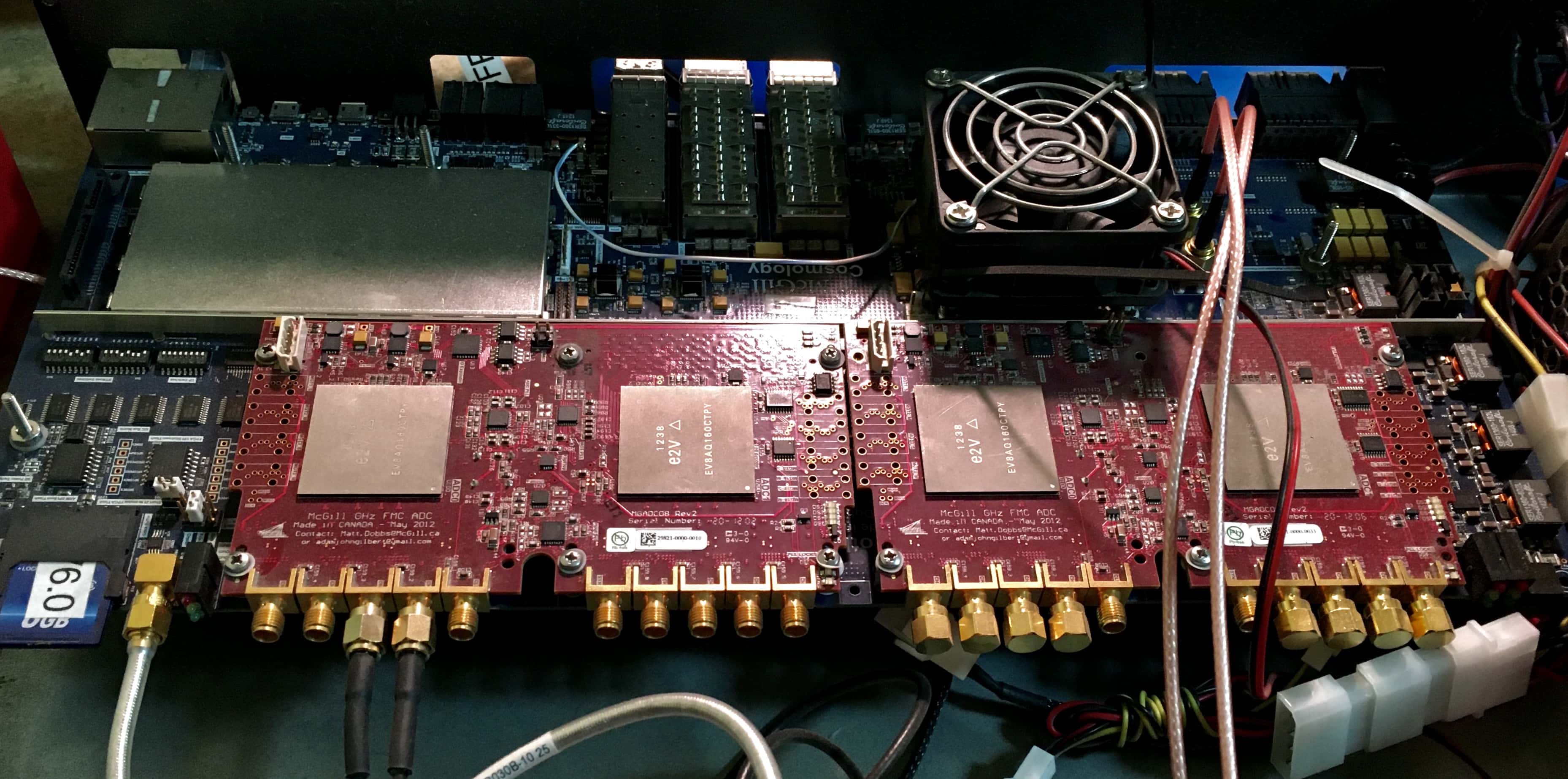}
    \caption{ICE, FPGAs, and ARM processor in control room of the ARO 10-m telescope.}
    \label{fig:fpgas}
\end{figure}

The ICE system hardware \citep{2016JAI.....541005B} at the ARO 10-m telescope is made of two ADC data-acquisition daughter boards attached to the field-programmable gate array (FPGA), and an ARM processor running Linux which loads the firmware onto the FPGAs (see Figure \ref{fig:fpgas}) and provides monitoring functionality.
The ICE reads data from the ADCs at \SI{8}{\bit} with a sample rate of \SI{800}{\MSPS}, resulting in a data rate of \SI{6.4}{\giga\bit\per\s} for each of the ADCs.
The data products from the ICE (after PFB and quantization) go to the main recording node on site, where the data are carried through two QSFP+ connectors on the motherboard which then become eight connections with SFP+ ports.
The total possible data rate on the recording node is therefore \SI{51.2}{\giga\bit\per\s}, but the actual data rate used is \SI{6.4}{\giga\bit\per\s} from 2 of the 16 ADC inputs (in the \SI{51.2}{\giga\bit\per\s}).

The main difference between the signal chains of CHIME/FRB and the ARO 10-m telescope is the fact that CHIME/FRB uses a GPS disciplined crystal oscillator for its \SI{10}{\mega\hertz} clock, thereby keeping the clock in-sync with GPS on long timescales. Separately we digitize a maser signal (DRAO maser), in order to correct for timing variations of the GPS clock (on short timescales), i.e., CHIME/FRB has effectively a maser-precision. This combines the long term absolute accuracy the GPS clock with the short timescale relative precision of the maser \citep{2021arXiv211000576M}.

On the other hand, the \SI{10}{\mega\hertz} clock at ARO is provided by a hydrogen maser (which provides tempo), and the GPS unit only provides an initial reference GPS timestamp via an IRIG-B connection to the ICE system. The clock is therefore free running off the maser and not adjusted to keep time with GPS after the start of an acquisition.
In VLBI, clock stability is crucial to understand and localize sources (due to its geometric relation, see Section \ref{sec:vlbi_between_the_aro_10-m_and_chime}), and its reliability is measured with the Allan variance (standard measure of frequency stability in clocks).

As in CHIME, the data processing system used at the ARO 10-m telescope is \texttt{kotekan}\footnote{The \texttt{kotekan} software repository: \url{https://github.com/kotekan/kotekan}.}, which is run in the recording node. This is a framework for assigning blocks of processing components. The \texttt{kotekan} software works same as in CHIME but its configuration appropriate for single-antenna recording.

There is currently no real-time processing of baseband data at the ARO 10-m telescope; instead, all data are stored in a set of \num{10} drives of \SI{11}{\tera\byte} each, giving a buffer with a theoretical capacity  of \SI{30}{\hour}. In reality, this time period is lower due to the limitation of the data writing to disk, the space used in the drives, and the constantly-running disk-cleaner utility.
Accounting for this, the actual buffer is roughly $\SI{\sim24}{\hour}$. The recording depends on: maintenance, power outages, and other down times.
This window is more than enough to establish a connection with CHIME/FRB and a potential triggering signal. The data are recorded to the set of hard drives simultaneously for fast and efficient storing and processing on the recording node.
Data that have been on the disks for longer than the buffer time are eventually deleted by the disk-cleaner (which is a simple script used to remove data after a certain disk usage has been reached), unless the system is stopped manually or data have been hard-linked\footnote{Hard-linking refers to a data link, or in effect a copy, that frees random-access memory (from a normal copying process) and allows the data to be accessed even after the original has been erased (original data bits are preserved).} by a CHIME/FRB triggering event.

Important elements in this chain such as the FPGAs, \texttt{kotekan}, disk-cleaner, and hard-linker are automatically serviced on the recording node and constantly monitored by users.

\begin{figure}[t]
    \centering
    \includegraphics[width=.4\textwidth]{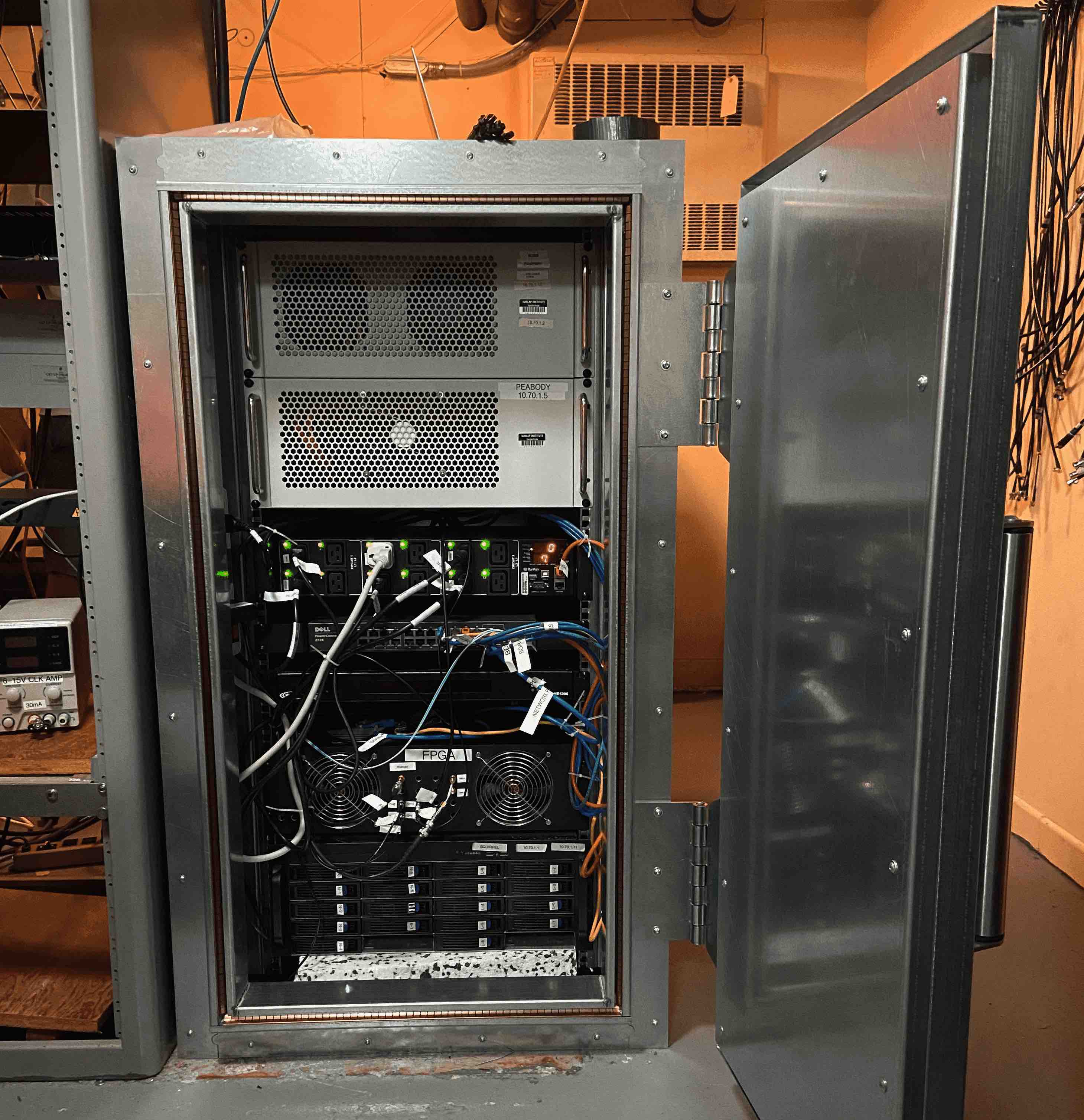}
    \caption{RF-shielded rack located in the main control room at ARO. The rack includes (from top to bottom): controller node, analysis node, PDU, main switch, GPS unit, ICE/FPGAs, and recording node.}
    \label{fig:rack}
\end{figure}

\begin{figure}[t]
    \centering
    \includegraphics{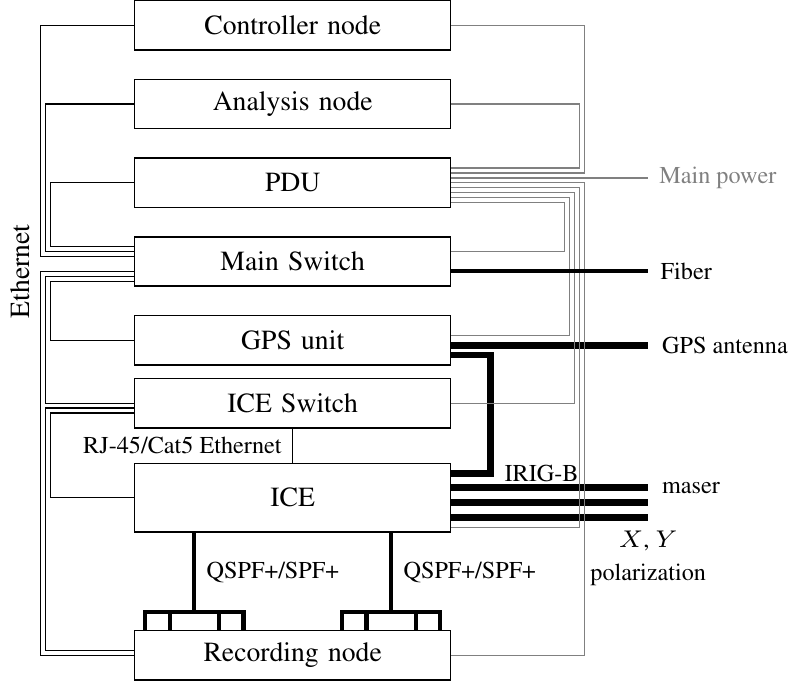}
    \caption{Digital chain connections diagram inside the RF-shielded rack. Thin black lines represent Ethernet connections, thin gray lines are power connections, thick lines are high speed links (RJ-45/Cat5 Ethernet and QSFP+/SFP+), and very thick lines are analog connections (SMA connectors). Terminal cables on the right side are connections that go to the exterior of the RF-shielded rack. This figure represents the schematics of Figure \ref{fig:rack}.}
    \label{fig:rack_digital}
\end{figure}

Other components of the digital chain are machines: controller node, which is in charge of the intensity stream from \texttt{kotekan} and receiving trigger signals from CHIME; and analysis node, the main machine used to check data and early science results before data are transferred to a common location. To minimize the RFI from different components in the analog chain, all systems described in this section are enclosed in an RF-shielded rack; see Figures \ref{fig:rack} and \ref{fig:rack_digital}.

Other network and power elements are the Main Switch and the Power Distribution Unit (PDU).

Site communication is done via Xplornet satellite-based Internet.
The Internet connection varies with weather with an average upload/download of \SI{4.5}{\mega\bit\per\s}.
Nevertheless, we only require checks on the telescope through Secure Shell (\texttt{ssh}), and once a day pulses (potential FRBs or calibrators) are transferred over to the Compute Canada system SciNet \citep{10.1145/3332186.3332195}. This is a common location where the cross-correlation process between the DRAO and ARO sites is carried out.

\subsection{Triggering from CHIME}
\label{sec:triggering_from_chime}

\begin{figure*}
    \centering
    \includegraphics{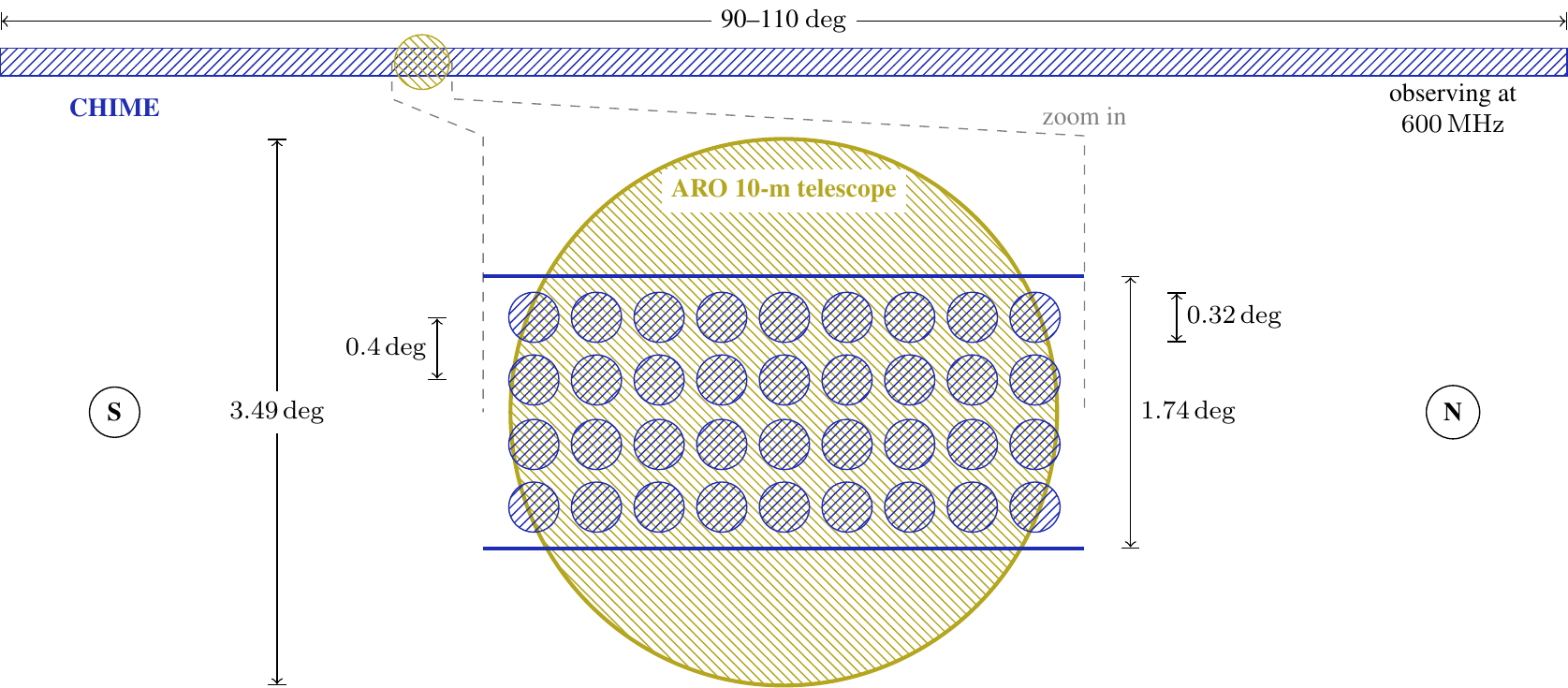}
    \caption{CHIME and ARO 10-m telescope FoV comparison. Figure is to scale comparing at an observing frequency of \SI{600}{\mega\hertz}. The zoomed-in region shows single CHIME/FRB beams of size \SI{0.32}{\deg} with a separation between them of \SI{0.4}{\deg}. The ARO 10-m telescope beam is located on the southern section of the CHIME beam declination \SI[parse-numbers=false]{+22}{\deg}.
    The CHIME/FRB beamformed beams change in shape with zenith angle \citep{2017arXiv170204728N}.}
    \label{fig:fov}
\end{figure*}

The ARO 10-m telescope data are not searched independently for astrophysical transients. Data are only saved upon receipt of a trigger from CHIME/FRB. Only FRBs that are detected in the shared FoV between CHIME/FRB and the ARO 10-m telescope cause a trigger to the ARO site.
The overlapped FoV is centered at declination \SI[parse-numbers=false]{+22}{\deg} and only beams in the same sky location as the 10-m are able to trigger events (roughly \num{36} beams at \SI{600}{\mega\hertz}, see Figure \ref{fig:fov}). Baseband data for processed events are then triggered at CHIME (within the same CHIME/FRB pipeline as normal events, see \citeauthor{2018ApJ...863...48C} \citeyear{2018ApJ...863...48C}) and a trigger is sent to ARO, which activates hard-link storage and backs up the recently recorded data. For calibration and delay-reference purposes, pulses from PSR B0531+21 (the Crab pulsar) are also dumped daily in the same way as an FRB.

CHIME/FRB detection happens in beamformed beams \citep{2017arXiv170204728N}, which then alert the ARO 10-m telescope with a triggering signal. Figure \ref{fig:fov} shows an up-to-scale comparison of CHIME/FRB beams and the single beam from the ARO 10-m telescope. With this current configuration, we expect to observe \SIrange{\sim4}{7}{\FRBs\per\month} (estimated from CHIME/FRB rates without considering down times). However, only the brightest pulses are detected in cross-correlation.

The signal-to-noise ratio is only computed at CHIME, this is because of the following reasons:
\begin{itemize}
    \item PSR B0531+21 pulses are observed through the Crab Nebula which will increase the system temperature (irrelevant for FRBs). 
    \item Some pulses are not visible in intensity at the ARO 10-m telescope, but they cross-correlate.
    \item The ARO 10-m telescope pointing is only known within \SI{\sim1}{\deg}, meaning that some of the detected pulses could fall in a side lobe.
    \item System temperature and beam model are unknown for the ARO 10-m telescope.
\end{itemize}

A single pulse from PSR B0531+21 (J0534+2200) \citep{2015MNRAS.446..857L} is recorded daily at the 10-m and at CHIME. These are giant pulses (GPs), with a  flux density of order \SIrange{1}{10}{\kilo\jansky}, of dispersion sweep \SI{1.1}{\s} (dispersion measure approximately \SI{56}{\parsec\per\centi\m\cubed}) at \SIrange{400}{800}{\mega\hertz}, and unknown Time of Arrival (ToA). Only a single pulse per day is triggered because of limitations of the CHIME/FRB system and the short time over which the two FoV overlap (roughly \SI{\sim10}{\minute}).
The recording system in CHIME/FRB is limited to \SI{100}{\milli\s} duration snapshots and has a trigger cooldown time of \SI{\sim5}{\minute}.

\begin{figure}[t]
    \centering
    \includegraphics{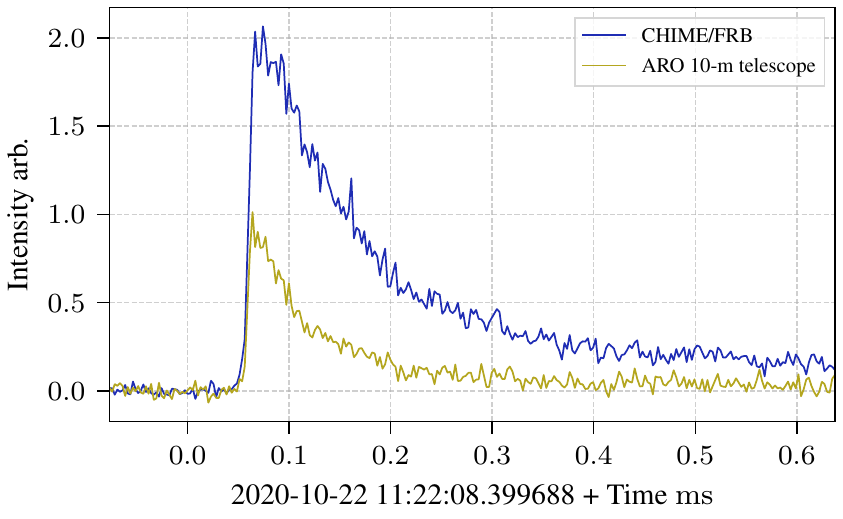}
    \caption{PSR B0531+21 pulse triggered at CHIME/FRB and also selected at ARO 10-m telescope. The intensity data are the power of each polarization combined per telescope and averaged over frequency channels. Data have been scaled in order to have the same level of the off-pulse rms. Both pulses have been coherently dedispersed to \SI{56.7558}{\parsec\per\centi\m\cubed}. The timestamp corresponds to the ToA measured at CHIME with an uncertainty of \SI{2.56}{\micro\s}. The ARO 10-m data will have a different start time adjusted to the geometric delay and clock errors.}
    \label{fig:intensity_comparison}
\end{figure}

The simultaneous baseband dumps are used on a regular basis to check data quality and clock stability, study the ionosphere, and serve as a potential calibrator for an FRB event before or after its recording.
The single pulse is selected via a signal-to-noise ratio threshold at CHIME of $\text{SNR}_\text{CHIME}\geq30$ and then triggered at the ARO 10-m telescope with the selected CHIME/FRB timestamp $\SI{\pm2.45}{\milli\s}$ (geometric delay due to baseline $b$). An example of a pulse observed simultaneously at CHIME/FRB and ARO 10-m telescope is in Figure \ref{fig:intensity_comparison}.

Besides standard triggering using the CHIME/FRB backend, CHIME is equipped with a VLBI tracking beam that can digitally point and save baseband data to disk. Observations with the CHIME VLBI tracking beam are a continuous stream of data (similar to the ARO 10-m telescope), and the CHIME VLBI tracking beam doesn't receive triggers. The CHIME VLBI tracking beam was used to observe PSR B0531+21 in continuum mode (i.e., several minutes recorded) because the CHIME/FRB trigger system could not record raw voltage data from multiple pulses on such short time scales. These observations were used to study the clock stability over a single day in Section \ref{sec:clock_stability_single_day}.

\subsubsection{Baseband data}
\label{sec:baseband_data}

\begin{figure}[t]
    \centering
    \includegraphics{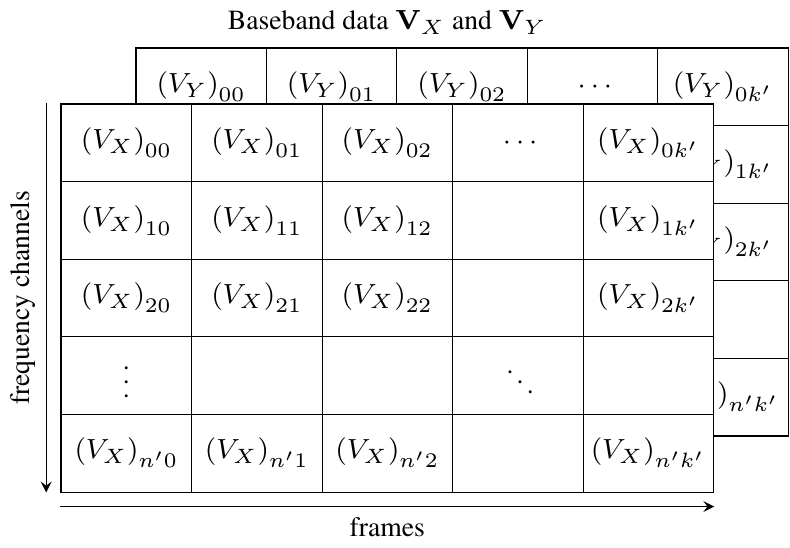}
    \caption{Visual representation of CHIME and ARO 10-m telescope baseband data, $\mathbf{V}$. For short notation $k'=K-1$ and $n'=N-1$ were used.
    Each polarization has $N\times K$ complex numbers, with $N=1024$ frequency channels and $K$ the number of frames. The quantities $\del{V_X}_{nk}$ and $\del{V_Y}_{nk}$ (with $k=0, \dotso, K-1$ and $n=0,\dotso, N-1$) are complex numbers \texttt{np.cdouble}. Each element $\del{V_X}_{nk}$ or $\del{V_Y}_{nk}$ has a GPS timestamp associated with a time bin of \SI{2.56}{\micro\s} called frame.}
    \label{fig:baseband}
\end{figure}

As mentioned in Section \ref{sec:digital_processing}, both CHIME and the ARO 10-m telescope return a similar type of baseband data as a product. The complex voltage is stored in an array of $N=1024$ frequency channels, $K$ time bin width of \SI{2.56}{\micro\s}, each of which is called a \textit{frame}, and dual linear polarizations (in CHIME North-South $Y$ and East-West $X$ polarizations), hereafter called \textit{baseband data} (after passing through the PFB). Figure \ref{fig:baseband} shows a representation of the baseband data, where the elements $\del{V_X}_{nk}$ and $\del{V_Y}_{nk}$ with $k=0, \dotso, K-1$ and $n=0,\dotso, N-1$; are complex numbers which are part of the $X$ and $Y$ polarizations. Baseband data will be represented as $\mathbf{V}$:

\begin{gather}
   \mathbf{V}_X\sbr{n, k} = \del{V_X}_{nk}\in\mathbb{C}^{N\times K} \quad\text{or} \\ \mathbf{V}_Y\sbr{n, k}=\del{V_Y}_{nk}\in\mathbb{C}^{N\times K}, \nonumber
\end{gather}
depending on their polarization. The notation $\sbr{n, k}$ refers to discrete dependency on frequency channels ($n$) and frames ($k$).
The first channel $n=0$ has a center frequency of \SI{800}{\mega\hertz} and the last channel $n=1023$ is \SI{400.390625}{\mega\hertz}.

At the ARO 10-m telescope and the CHIME VLBI tracking beam, data are stored to disk as a continuum, i.e., column $n=0$ has the same start time for all its elements. In contrast, the CHIME/FRB backend sends callback detections from intensity data and clips around baseband data on buffer; the clipping returns a block of baseband data where frequency channels have different start times and follow the dispersion of a pulsar or FRB.

A higher time resolution than \SI{2.56}{\micro\s} can be achieved by inverting the PFB (raw ADC data); this is a complicated procedure, but in principle is feasible. Alternatively, this can be achieved by studying the phase information in the complex number frames of the baseband data. In this analysis we will only use the latter, baseband data space. For a more detailed description of the baseband and raw ADC data properties see Appendix \ref{sec:visibility_and_cross-correlation_in_baseband_space}.

\section{VLBI between the ARO 10-m telescope and CHIME}
\label{sec:vlbi_between_the_aro_10-m_and_chime}

We are interested in finding the cross-correlation that exists between the same wavefront detected on different sites. Once the wavefront reaches CHIME  and the ARO 10-m telescope (hereafter site $A$ and site $B$), it is digitized and transformed to baseband data. 
The telescopes $A$ and $B$ will observe delayed copies of the same signals. The non-dispersive delays contain localization information. But also other sources of non-dispersive delay will exist such as the troposphere and the clock error between stations. In addition, the dispersive delays given by the ionosphere and instrumental effects must be properly calibrated.

Then starting from $\del{\mathbf{V}_{P_A}, \mathbf{V}_{P_B}}$ collected from the event at both sites\footnote{Here, the polarizations $P_A$ and $P_B$ can be either a linear $(X, Y)$ or circular $(R, L)$ basis.}, we will define the sub-frame cross-correlation $\rho_{A, B}^\text{sf}(\tau)$, a function of delay $\tau$. The quantity $\rho_{A, B}^\text{sf}(\tau)$ (mathematically defined in Section \ref{sec:correlation_with_baseband_data}) is only computed at a sub-frame level and it tells the time difference (lag) relative to time measured at $A$ and $B$. 

\begin{figure}[t]
    \centering
    \includegraphics{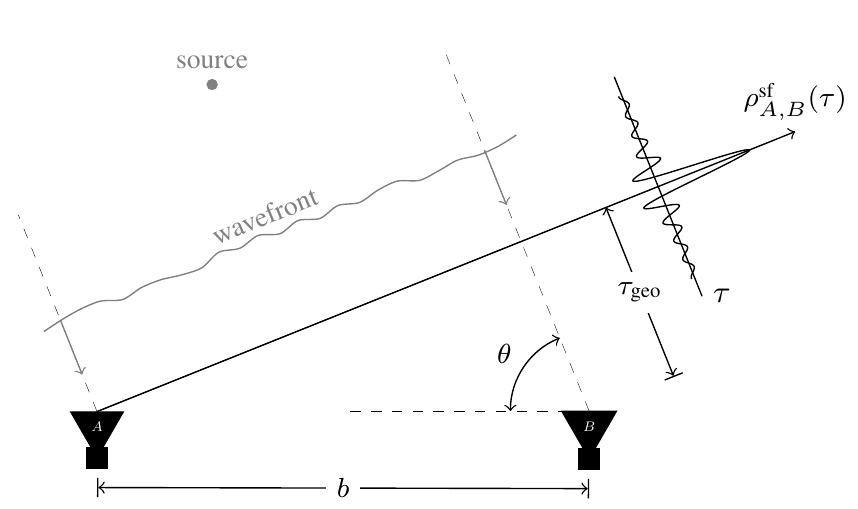}
    \caption{VLBI of a distant source of radio noise. The sub-frame cross-correlation function $\rho_{A, B}^\text{sf}(\tau)$ will be maximum when the delay $\tau_\text{geo}$ is applied. The variable $b$ represents the baseline (same as in Figure \ref{fig:vlbi_projection}) Euclidian distance between sites $A$ and $B$, and $\theta$ is the baseline angle which is function of RA and DEC, $\theta=\theta(\upalpha,\updelta)$.
    Notice that the wavefront is only an approximation of a plane wave; the atmosphere and interstellar medium will modify its structure (see total delay Eq.~\ref{eq:delays}).}
    \label{fig:VLBI_basics}
\end{figure}

Figure \ref{fig:VLBI_basics} shows the basic components from the VLBI system. We are interested in finding the localization given by the baseline angle $\theta$ (angle between the baseline vector and the source vector), which is function of the RA and DEC $\theta=\theta(\upalpha,\updelta)$ of the source, and in our case most of the contribution will be given by the RA (Figure \ref{fig:vlbi_projection}). The baseline angle is expressed simply as:

\begin{equation}
    \theta = \arccos\del{\frac{\tau_\text{geo} c}{b}},
    \label{eq:baseline_angle}
\end{equation}

with $c$ the speed of light, $b$ the baseline ($b\approx\SI{3000}{\kilo\m}$), and $\tau_\text{geo}$ the geometric delay. 
%

For the ARO 10-m telescope testbed, calibrators are pulsars (in particular, PSR B0531+21 pulses), since pulsars are easier to tell apart from the background as compared to standard VLBI (steady source) calibrators. Pulsars, in addition, are sufficiently compact and with no confusion (time-domain separation from background) at our frequencies.

The CHIME and ARO 10-m telescope stations have system-equivalent flux densities (SEFD; $S_\text{sys}$) of \SI{\sim40}{\kilo\jansky} and \SI{\sim1.7}{\kilo\jansky}. When Taurus A is in the beam (e.g., during a Crab GP trigger), the SEFD at the ARO 10-m telescope increases to \SI{1}{\kilo\jansky} ($S_\text{Taurus A}\approx\SI{1}{\kilo\jansky}$; \citealt{2017ApJS..230....7P}).
The CHIME observation will be completely dominated by the Taurus A flux \citep{2004ApJ...612..375C}. VLBI observations of the Crab pulsar baseline noise level will not be discussed, since the nebula is not correlated (\citealt{2017ARep...61..178P}; nebula determines the noise level and does not lead to a correlation), with the exception of Crab's GPs. In addition, Crab GPs can easily exceed the Taurus A emission with fluxes above $\SI{1}{\kilo\jansky}$ \citep{2021ApJ...920...38B,2021MNRAS.508.1947T}, i.e., $S_\text{sys}\approx S_\text{GPs}$ (Section \ref{sec:triggering_from_chime}; since only high enough signal-to-noise ratio pulses are correlated).

Earlier tests were performed at CHIME frequencies with steady source calibrators in VLBI but correlations were not possible to achieve. Steady source calibrators are known to be sufficiently unresolved at high frequencies and narrow bandwidths, which is not the case at CHIME and ARO 10-m telescope.
Nevertheless, the future CHIME/FRB Outrigger project is taking steady-source VLBI calibrators into consideration as LOFAR has been successful in doing so at low frequencies \citep{2015A&A...574A..73M}. The cross-correlation using pulsars is in principle similar to that for an FRB (with the exception of longer dispersion times, as will be discussed in detail in Section \ref{sec:geometric_delay_and_high_dm}).

We must find and isolate the geometric delay $\tau_\text{geo}$ from the total (observed) delay $\tau_\text{total}$. We can then decompose it into its different components:

\begin{equation}
    \tau_\text{total}(t, \nu) = \tau_\text{clock} + \tau_\text{geo}(t, \theta) + \tau_\text{iono}(\nu) + \tau_\text{inst}(\nu) + \xi. \label{eq:delays}
\end{equation}
In Eq.~\eqref{eq:delays}, left to right, the delays are: total, clock error (constant), geometrical (baseline angle and time-dependent, which also includes tropospheric delay), ionospheric (frequency-dependent), instrumentation (frequency-dependent), and noise errors.

A general introduction to VLBI and how to treat and remove delays from Eq.~\eqref{eq:delays} is described by \citet{2014ARA&A..52..339R}.

\subsection{Correlation with baseband data}
\label{sec:correlation_with_baseband_data}

In a traditional XF correlator framework \citep{2016arXiv160703579P}, raw ADC sampled data are cross-correlated with efficient Fast Fourier Transform (FFT) algorithms without any hierarchical separation of time or frequency scales, with the peak of the one-dimensional cross-correlation function yielding a measurement of the geometric delay, $\tau_\text{geo}$.
One drawback of this approach is that the problem is further complicated by the introduction of nanosecond-scale systematic drifts. All of these drifts change on minute- to day-long timescales, and they exhibit different frequency dependencies.
It is desirable to correlate our baseband data in a more hierarchical manner so we can characterize and separate these different shifts by their differing frequency dependence and time dependence. We break down the problem into two parts: first we determine the integer delay, measured in multiples of the baseband period of \SI{2.56}{\micro\s} (number of $k$ to shift), followed by the fractional (or sub-frame) part of the delay (phase correction), according to

\begin{equation}
    \frac{\tau_\text{total}}{\SI{2.56}{\micro\s}} = \cbr{\frac{\tau_\text{total}}{\SI{2.56}{\micro\s}}} + \left\lfloor\frac{\tau_\text{total}}{\SI{2.56}{\micro\s}}\right\rfloor,
\end{equation}
where $\lfloor.\rfloor$ is the integer part of a frame, and $\cbr{.}$ the fractional part. At the integer frame level, the largest delays to compensate for are the geometric and clock delays, i.e.

\begin{equation}
    k_\text{shift} = \left\lfloor\frac{\tau_{\text{geo}_0} + \tau_{\text{clock}_0}}{\SI{2.56}{\micro\s}}\right\rfloor. \label{eq:k_shift}
\end{equation}
Where $\tau_{\text{geo}_0}$ is given by a prior source localization which is accurate at the arcminute level (Section \ref{sec:geometric_correction}), and $\tau_{\text{clock}_0}$ is a prior measurement of the constant delay.

\begin{figure}[t]
    \centering
    \includegraphics{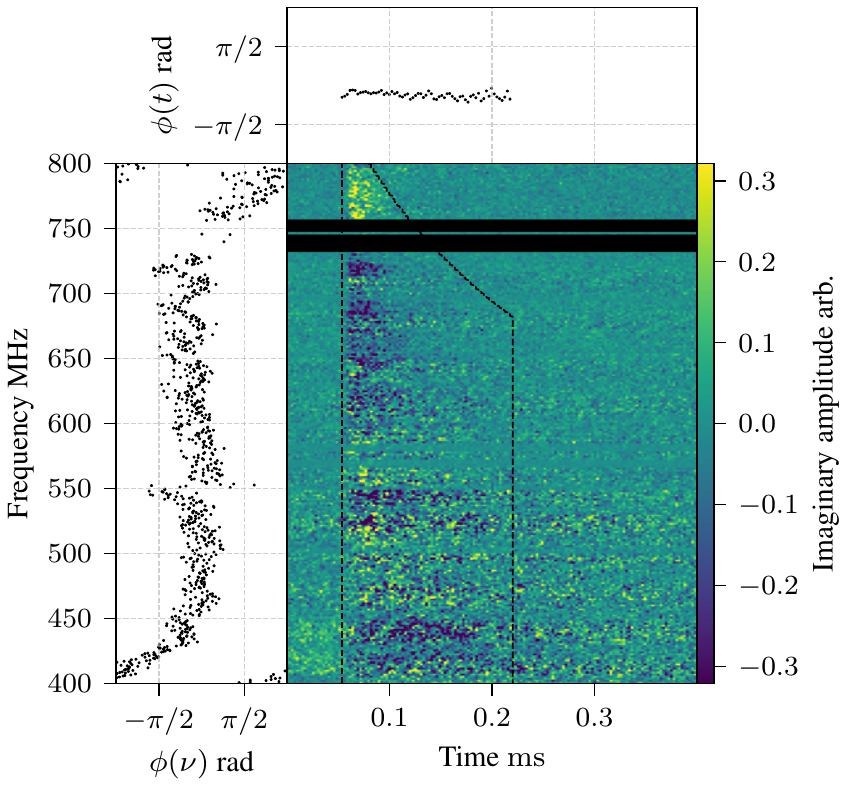}
    \caption{Imaginary part of visibility, $\text{Im}\sbr{\boldsymbol{\mathcal{V}}_{Y_AX_B}}$. Left and top panels are the integrated (summed) visibility angle over time and over frequency. The integration is done along the strongest section of the pulse width $\del{t_\text{w}}$ taking into consideration the scattering tail profile $\propto\nu^{-4}$ (curved black dashed line).
    For these strong pulses (same from Figure \ref{fig:intensity_comparison}) its phase wrapping is close to \SI{0}{\radian}. Other conjugates of polarization pairs can be seen in Appendix \ref{sec:visibility_and_cross-correlation_in_baseband_space}, Figure \ref{fig:correlation_matrix}. Black horizontal lines are dead frequency channels.}
    \label{fig:phase_baseband}
\end{figure}

If we have measured the frame delay correctly, $k_\text{shift}$, the visibility matrix can be constructed:

\begin{align}
    \boldsymbol{\mathcal{V}}_{P_AP_B}\sbr{n, k} &= \mathbf{V}_{P_A}\sbr{n, k-k_\text{shift}} \overline{\mathbf{V}_{P_B}\sbr{n, k}} \nonumber \\&= \del{\mathcal{V}_{P_AP_B}}_{nk}\in\mathbb{C}^{N\times K}, \label{eq:visibility}
\end{align}
with $\boldsymbol{\mathcal{V}}_{P_AP_B}$ the matrix of frames and frequency channels with $\del{\mathcal{V}_{P_AP_B}}_{nk}$ elements, which is simply an array element-by-element multiplication. The complex conjugate of all elements in matrix $\mathbf{V}$ is expressed as $\overline{\mathbf{V}}$.
Equivalently we can rewrite Eq.~\eqref{eq:visibility} in terms of its phase:

\begin{equation}
    \boldsymbol{\mathcal{V}}_{P_AP_B} = \enVert{\boldsymbol{\mathcal{V}}_{P_AP_B}}\mathrm{e}^{\mathrm{i}\boldsymbol{\phi}}, \quad \boldsymbol{\phi}=\text{Arg}\sbr{\boldsymbol{\mathcal{V}}_{P_AP_B}}, \label{eq:vis_phi}
\end{equation}
with $\boldsymbol{\phi}$ a matrix with each element $\phi_{nk}$ (frames and frequency channels), and \enVert{.} the norm of each complex element.
The sub-frame delay of Eq.~\eqref{eq:visibility} shows up in the phase of each complex number. One approach is to take the Fourier Transform of $\boldsymbol{\mathcal{V}}_{P_AP_B}$ over the frequency axis ($n$).
This is approximately equivalent to lag-correlating our baseband data in many small \SI{2.56}{\micro\s} integrations, ignoring spectral leakage effects. If the delay were purely geometric, finding a peak in sub-frame cross-correlation, as a function of $\tau$ would perfectly determine the sub-frame part of the delay. However this approach does not allow us to separate out the different contributions to the delay, as we discussed earlier.
Instead, we sum (integrate) over frames, measuring the phase as a function of frequency channel, as shown in Figure \ref{fig:phase_baseband}, left panel. Then the \textit{integrated visibility} is defined as:

\begin{equation}
    \left<\boldsymbol{\mathcal{V}}_{P_AP_B}\right>_t\sbr{n} = \sum_{k'} \del{\mathcal{V}_{P_AP_B}}_{nk'},
    \label{eq:visibility_integrated}
\end{equation}
integrated over a time $t=k'\times\SI{2.56}{\micro\s}$ proportional to a number of frames, $k'$. We are left with a sequence of $N = 1024$ complex visibilities which contain all the sub-frame delays.

We now define the sub-frame cross-correlation function in terms of the integrated visibility of baseband data as:

\begin{align}
    \boldsymbol{\rho}^\text{sf}_{A, B}\sbr{u} &= {\sum^{N-1}_{n=0} \sum_{k'}\del{\mathcal{V}_{P_AP_B}}_{nk'} \mathrm{e}^{-\frac{\mathrm{i}}{N}2\pi un}} \nonumber \\
    &={\text{FFT}\sbr{\left<\boldsymbol{\mathcal{V}}_{P_AP_B}\right>_t}}, \label{eq:rhosf_visibility}
\end{align}
where the norm of the FFT over the frequency axis with $n$ channels and it returns the cross-correlation strength as a function of lag with $u$ data points, cross-correlation strength: $\enVert{\boldsymbol{\rho}^\text{sf}_{A, B}}$. Notice that the maximum value of the cross-correlation strength is:

\begin{equation}
    \tau_u = \argmax_{\tau_u\in\sbr{\SI{-1.28}{\micro\s}, \SI{1.28}{\micro\s}}}\enVert{\boldsymbol{\rho}^\text{sf}_{A, B}\sbr{u}}, \label{eq:lag-correlation}
\end{equation}
with $\tau_u$ the \textit{lag-correlation} or first estimate of the constant delay that exists between the two data sets.
To separate the different contributions to the total sub-frame delay, we must fringe-fit the visibilities $\left<\boldsymbol{\mathcal{V}}_{P_AP_B}\right>_t$ from several sources to a delay model (dispersive and non-dispersive) using the differing frequency and time dependencies.

\subsection{Geometric correction}
\label{sec:geometric_correction}
Due to the large value of the geometric delay (several frames) and its considerable variation over time $\tau_\text{geo}(t, \theta)$, an initial guess for $\tau_\text{geo}$ is required. The geometric delay model, $\tau_{\text{geo}_0}$, is used to re-align phase and frames of baseband data for correlation, but since it depends on the precision of the initial localization guess given by CHIME (or the precision of the baseline), a residual will be left to compute $\tilde{\tau}_\text{geo}$ here, where:

\begin{equation}
    \tilde{\tau}_\text{geo} = \tau_\text{geo} - \tau_{\text{geo}_0}.
    \label{eq:tau_geo_residual}
\end{equation}
This residual calculation will be done in the wideband fringe fitting algorithm (Section \ref{sec:wideband_fringe_fitting}).

Each baseband frame is tagged with a local atomic clock (or close to atomic performance). As such, we need to know the proper time delay between the arrival of the wavefront at one antenna and the arrival at the second. Computing this is complicated by the reality that the object under observation is only stationary in a frame (of reference) in which the antennas are moving. For short baselines, this motion is negligible compared to the time it takes light to cross the baseline, but for VLBI it can become significant.
Geometric delays will be computed using \texttt{difxcalc11} \citep{2016ivs..conf..187G}, where by giving an observation time, sky coordinates, and Earth locations a $\tau_{\text{geo}_0}$ model can be calculated. The software first computes the exact site locations, taking into account crust deformations due to a high-order tide model (and if ocean coefficients are available it will include those effects). Once it has the baseline, it applies the Consensus \citep{eubanks_proceedings_1991} relativity model, which accounts for the baseline motion in the solar system barycenter frame and for gravitational and special time dilation effects.
Other additional delays are added afterwards such as the troposphere refractive delay of the source in each site. Geometric delay models are intensively studied in geodetic VLBI and tend to be described by the same models as in astrometry VLBI \citep{2020PASA...37...50T}.
The model at the CHIME and ARO 10-m telescope baseline using \texttt{difxcalc11} is able to achieve picosecond precision \citep{1991AJ....101.2306S}, which is more than sufficient for our purposes.

\subsection{Coherent delay-referencing}
\label{sec:coherent_delay-referencing}

In order to localize a target (FRB) given a reference (calibrator) we use \textit{delay-referencing}. All astrometric information will be contained only in the phase of the two set of visibilities, and by computing their phase difference a good estimate of their sky angular difference ($\Delta\theta$) can be computed. Other frequency dependent effects such as the instrumentation phase error (given by $\tau_\text{inst}(\nu)$) will also cancel off.
We then calculate the angle of the visibility ratio, where in the ideal case $\tau_\text{total}=\tau_\text{geo}$ the ratio becomes:

\begin{equation}
    \Delta\phi_\text{geo} = \phi^\text{t}_\text{geo} - \phi^\text{r}_\text{geo} = 2\pi b\frac{\nu}{c}\left(\cos\theta_\text{t} - \cos\theta_\text{r}\right), \label{eq:phase_ref}
\end{equation}
where $\theta_\text{t}$ is the baseline angle of the target, $\theta_\text{r}$ the baseline angle of the reference, and $\phi$ represents the angle of the complex visibility.
Notice that to reach Eq.~\eqref{eq:phase_ref} we need to first account for all terms from Eq.~\eqref{eq:delays} in the phase $\boldsymbol{\phi}$ (delays smaller than a frame) and by finding the right alignment of the two baseband data matrices, $k_\text{shift}$, at the time to form $\boldsymbol{\mathcal{V}}_{P_AP_B}$. In practice, the difference $\Delta\phi_\text{geo}$ can be related to a sky angular difference $\Delta\theta$ which will have the reference and target locations and it will be a function of $(\upalpha, \updelta)$.
The steps and assumptions to reach Eq.~\eqref{eq:phase_ref} are in Sections \ref{sec:wideband_fringe_fitting} and \ref{sec:towards_localization}. Finally in section \ref{sec:discussion} we will discuss the limitations of the method.

\subsection{Wideband Fringe Fitting}
\label{sec:wideband_fringe_fitting}
Fringe Fitting is the method used to find a likelihood distribution that represents the most accurate parameters for geometric delay and time degeneracies in order to maximize the cross-correlation strength (or equivalently minimize $\chi^2$).
As mentioned earlier, the ARO 10-m telescope site is only intended as a coherent delay-reference telescope, hence we will not take into consideration other source properties for this analysis.
The fringe fitting algorithm will only be performed over the visibility as a function of frequency and normalizing by a strong (calibrator) visibility, as compared to the classical approach done at higher observational frequencies and narrow bandwidths (where delay and delay-rate may be slowly varying quantities and can be computed with a two dimensional Fourier Transform, see \citeauthor{1983AJ.....88..688S} \citeyear{1983AJ.....88..688S}).

\begin{figure}[t]
    \centering
    \includegraphics{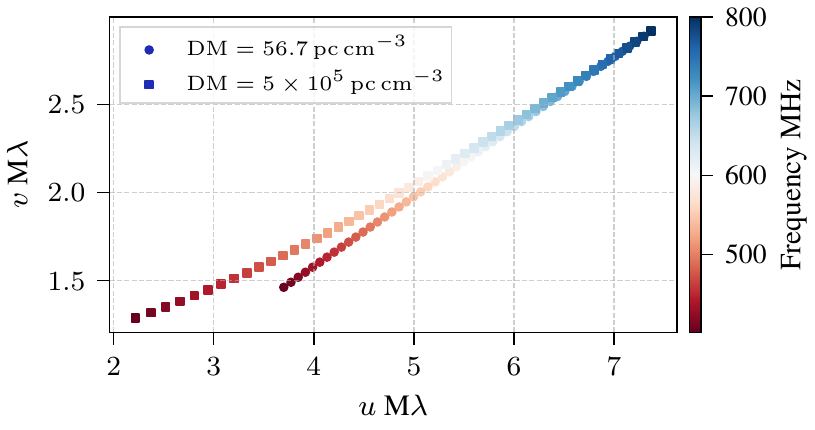}
    \caption{Dispersion effect over the \SIrange{400}{800}{\mega\hertz} band in the $uv$-plane. Two simulated pulses in the CHIME and ARO 10-m telescope baseline were simulated, with DM \SI{56.7}{\parsec\per\centi\m\cubed} (circle) and \SI{5e5}{\parsec\per\centi\m\cubed} (square). The latter is an exaggerated simulation to show the curving track of the $uv$-plane due to the frequency-dependent pulse ToA. The computation of the $(u, v)$ coordinates depends on the source location (both cases located at PSR B0531+21 RA and DEC), Earth locations and ToA (same for both pulses at \SI{800}{\mega\hertz}). The $u$- and $v$-axis are in \num{1e6} wavelengths units. The $(u, v)$ points are color-coded by the frequency from low (red) to high (blue), in the CHIME observing bandwidth.}
    \label{fig:uv_plane}
\end{figure}

In a traditional VLBI experiment the coverage of the $uv$-plane is a radial line whose length is related to the telescope bandwidth $\frac{\Delta\nu}{\nu_\text{obs}}$. Here the track would show curvature due to the changing baseline projection while the pulse arrives as a function of frequency depending on its DM. In the $uv$-plane this would be represented as shown in Figure \ref{fig:uv_plane}. For a lower DM comparable to PSR B0531+21 the effect is linear but of an FRB it may not be. Figure \ref{fig:uv_plane} shows the Crab pulsar and an exaggerated simulation of an FRB in the $uv$-plane.

After an FRB (target) and its calibrator (reference) have been observed (i.e., baseband data have been acquired on sites $A$ and $B$) we can proceed with forming visibilities and fringe fitting.
First we independently form two (or more) visibilities (Eq.~\ref{eq:visibility_integrated}), the target and the reference. The steps to form a visibility are as follows:

\begin{enumerate}
    \item Select a single polarization (either linear or circular basis) from site $A$ and $B$, i.e., $\mathbf{V}_{P_A}$ and $\mathbf{V}_{P_B}$. If the orientation of linearly polarized feeds is known then one could take the expected pair that maximizes the cross-correlation (e.g., highly polarized source). In our baseline we do not have this information and it is a trial and error process.

    \item Compute geometric correction (using \texttt{difxcalc11}) for a given $(\upalpha,\updelta)$; timestamp on site $A$, $t_A$; and Earth locations of sites $A$ and $B$.
    The process is repeated for each frame of the baseband data, hence $\boldsymbol{\tau}_{\text{geo}_0}\sbr{k} = \del{\tau_{\text{geo}_0}}_k\in\mathbb{R}^{K}$\footnote{Notice that for $\text{DM}\gtrsim\SI{200}{\parsec\per\centi\m\cubed}$ and $b>\SI{3000}{\kilo\m}$, the geometric delay across the band may be larger than a single frame, and $\left\lfloor\frac{\tau_{\text{geo}}}{\SI{2.56}{\micro\s}}\right\rfloor$ (step 5) will not be constant at \SIrange{400}{800}{\mega\hertz} (see Section \ref{sec:geometric_delay_and_high_dm}).}.
    Then decompose $\boldsymbol{\tau}_{\text{geo}_0}$ into an integer part of frames and a time residual.
    Apply the fractional frame part of the geometric correction (or the entire value, since a value larger than the frame size will wrap around $\SI[parse-numbers=false]{2\pi}{\radian}$) to the phase of the baseband data in a single site,

    \begin{gather}
        \mathbf{V}_{P_A}^{(\text{geo})} = \mathbf{V}_{P_A}\text{exp}\sbr{-\mathrm{i}\boldsymbol{\phi}_{\text{geo}_0}},\label{eq:baseband_geo_corr} \\ \boldsymbol{\phi}_{\text{geo}_0}\sbr{n, k} = 2\pi\boldsymbol{\tau}^\intercal_{\text{geo}_0}\sbr{k}\boldsymbol{\nu}\sbr{n}, \label{eq:phase_geo_correction}
    \end{gather}
    with $\boldsymbol{\phi}_{\text{geo}_0}$ the geometric phase correction and $\boldsymbol{\nu}$ the vector of all $N=1024$ frequency channels (and elements $\nu_n$). Eq.~\eqref{eq:baseband_geo_corr} contains a matrix, $\mathbf{V}_{P_A}$, where its phase is rotated by the vector $\boldsymbol{\phi}_{\text{geo}_0}=\del{\phi_{\text{geo}_0}}_k\in\mathbb{R}^{N\times K}$ and returns a $N\times K$ dimensions.
    Equivalently this correction can be applied to site $B$ with a positive phase term.

    \item Coherently dedisperse \citep{1975MComP..14...55H} pulse to an optimum dispersion measure (DM) (which is the same for both polarizations, $\mathbf{V}_{P_A}^{(\text{geo})}$ and $\mathbf{V}_{P_B}$), either by optimizing by structure or signal-to-noise ratio. A precision of roughly \SI{1e-3}{\parsec\per\centi\m\cubed} is needed in this step.
    The correction needed for a frequency channel is given by the dispersion time:

    \begin{equation}
        t_\text{dd}(\nu) = k_\text{DM}\text{DM}\frac{1}{\nu^{2}},
        \label{eq:dispersive_time}
    \end{equation}
    with $k_\text{DM}=\num{1}/(\num{2.41e-4})$ \si{\s\mega\hertz\squared\per\parsec\centi\m\cubed} is the dispersion constant \citep[see][Chapter~4]{2012hpa..book.....L}. In practice we need an algorithm that de-smears frequency channels (intra-channel alignment) and lines up the pulse to a reference frequency (inter-channel alignment). In this case, the natural kernel for
    coherent dedispersion is:

    \begin{equation}
        \Delta\psi = k_\text{DM}\text{DM}f\del{\frac{1}{f_\text{r}}-\frac{1}{f}}^2,
        \label{eq:kernel_cdd}
    \end{equation}
    with $f_\text{r}$ the reference frequency, $f$ the frequency and $\Delta\psi$ the phase amplitude of the transfer function \citep[see][Chapter~5]{2012hpa..book.....L}. After the transfer function has been applied, timestamps will only be relevant in the referenced channel, since the pulse itself will move on the baseband frames.
    Note that the geometric correction needs to be applied prior to the coherent dedispersion, and this is only the case because $\tau_\text{geo}$ is time dependent. After coherent dedispersion, the geometric delay is a function of the frame number and the frequency channel. Most of the dispersion $t_\text{dd}$ will be removed in this operation leaving a small (but significant) fraction due to the ionospheric electron content ($\tau_\text{iono}$).

    \item Align pulses up to a frame precision $\left\lfloor\frac{\tau_\text{total}}{\SI{2.56}{\micro\s}}\right\rfloor$; in practice it is only the geometric and clock error delays, $k_\text{shift}$ in Eq.~\eqref{eq:k_shift},
    which is adding an integer frame delay to $\mathbf{V}_{P_A}^{(\text{geo})}$ with respect to $\mathbf{V}_{P_B}$.
    The value for $\tau_\text{clock}$ is a changing quantity, varying significantly over days\footnote{In our baseline of the order of \SI{0.1}{\micro\s\per\day}, see Section \ref{sec:clock_stability_multiple_day}.}, and depends on the rate of the two masers (of site $A$ and $B$). This value will change more than a frame over a timescale of weeks, but in principle it is a small and known value from previous correlations and an estimate can be chosen $\tau_{\text{clock}_0}$\footnote{Similarly as in Eq.~\eqref{eq:tau_geo_residual}, the clock delay can be expressed as:
    \begin{equation}
        \tilde{\tau}_\text{clock} = \tau_\text{clock} - \tau_{\text{clock}_0} \nonumber
    \end{equation}}.

    \item Form integrated visibility (same as in Eq.~\eqref{eq:visibility}), and complex sum along the pulse width $t_\text{w}$ (in the case of Figures \ref{fig:intensity_comparison} and \ref{fig:phase_baseband}, $t_\text{w}\approx\SI{0.4}{\milli\s}$) Eq.~\eqref{eq:visibility_integrated},

    \begin{align}
        \boldsymbol{\mathcal{V}}_{P_AP_B}^{\text{(geo)}}\sbr{n, k} &= \mathbf{V}_{P_A}^{\text{(geo)}}\sbr{n, k-k_\text{shift}} \overline{\mathbf{V}_{P_B}\sbr{n, k}} \nonumber \\
        \left<\boldsymbol{\mathcal{V}}_{P_AP_B}\right>_{t_\text{w}}\sbr{n} &= \sum_{k=t_0-\frac{1}{2}t_\text{w}}^{k=t_0 + \frac{1}{2}t_\text{w}} \del{\mathcal{V}_{P_AP_B}^{\text{(geo)}}}_{nk},\nonumber
    \end{align}
    with $t_0$ the pulse center, $t_0\pm \frac{1}{2}t_\text{w}$ is an integer number of frames, and $\left<\boldsymbol{\mathcal{V}}_{P_AP_B}\right>_{t_\text{w}}$ (from now on dropping (geo) superscript) will only depend on frequency channels.
    Note that if strong scattering is present, then we may need to adjust the pulse profile with a frequency-dependent broadening function to increase signal-to-noise ratio. Then each frequency channel will have a pulse profile as $t_\text{w}\propto\nu^{-4}$.
\end{enumerate}
After the two visibilities are formed, we can start fringe fitting. We first reference the source to the calibrator by making the ratio of visibilities (equivalent to Eq.~\eqref{eq:phase_ref}),

\begin{equation}
    \boldsymbol{\mathcal{V}}_\text{norm}\sbr{n} = \frac{\left<\boldsymbol{\mathcal{V}}^\text{t}_{P_AP_B}\right>_{t_\text{w}}}{\left<\boldsymbol{\mathcal{V}}^\text{r}_{P_AP_B}\right>_{t_\text{w}}},
    \label{eq:vis_ratio}
\end{equation}
with $\boldsymbol{\mathcal{V}}_\text{norm}$ the normalized (integrated) visibility (vector size $N$), and each element $\del{\mathcal{V}_\text{norm}}_n$ (i.e., the sum or integration of each frequency channel $n$).
Notice that a Fourier Transform over the frequency channels axis of $\boldsymbol{\mathcal{V}}_\text{norm}$ (Eq.~\ref{eq:vis_ratio}) represents the lag-correlation, discussed in Section \ref{sec:correlation_with_baseband_data}.

By taking the ratio in Eq.~\eqref{eq:vis_ratio} we are effectively computing Eq.~\eqref{eq:phase_ref}; the ratio will remove other delay terms such as frequency dependent instrumentation effects ($\tau_\text{inst}(\nu)$) and repetitive errors ($\xi$). By doing so we are just left with the residuals for $\tau_\text{iono}$, $\tau_\text{geo}$, and $\tau_\text{clock}$. We will define residuals as: $\delta\tau_\text{iono}$ ionospheric residual and $\delta\tau$ for all other constant residuals left.
The ratio will be most effective (or the residuals will be the smallest) when the source and calibrator are observed closer in time and in sky angle. The former is due to the fact that there will be less clock residual to correct, and the latter because there will be less delay contributions from the ionosphere.

Now by finding the residual terms we can optimize a likelihood function for the visibility model \citep{1999ASPC..180..335P},

\begin{equation}
    \mathcal{L} \propto\prod_{n=0}^{N-1} \text{exp}\left[-\frac{1}{2}\sigma_n^{-2}\enVert{\del{\mathcal{V}_\text{norm}}_n - F(\nu_n; \delta\tau, \delta\tau_\text{iono})}^2\right],\nonumber
\end{equation}
with $F$ the fringe model to be fitted and $\sigma_n^2$ the variance\footnote{The variance of a complex number is expressed as:
\begin{equation*}
    \sigma^2 \equiv \text{Var}\left[\boldsymbol{\mathcal{V}}\right] = \text{Var}\left[\text{Re}\left[\boldsymbol{\mathcal{V}}\right]\right] + \text{Var}\left[\text{Im}\left[\boldsymbol{\mathcal{V}}\right]\right],
\end{equation*}
with $\sigma^2\in\mathbb{R}^+$.} of the complex random variable $\boldsymbol{\mathcal{V}}_\text{norm}$, and $N$ the total number of frequency channels ($\nu_n$ channel $n$). In principle, the localization is only present in the phase of the visibility and not in its amplitude, and if errors are normally distributed then,
\begin{eqnarray}
    \chi^2 &=& \sum^{N-1}_{n=0}\frac{\enVert{\del{\mathcal{V}_\text{norm}}_n - F(\nu_n; \delta\tau, \delta\tau_\text{iono})}^2}{\sigma_n^2}\nonumber\\
    &=& \sum^{N-1}_{n=0}\frac{\enVert{\del{\mathcal{V}_\text{norm}}_n - \enVert{F(\nu_n)}\mathrm{e}^{\mathrm{i}\varphi(\nu_n; \delta\tau, \delta\tau_\text{iono})}}^2}{\sigma_n^2}\nonumber\\
     &=& \sum^{N-1}_{n=0}\frac{\enVert{\del{\mathcal{V}_\text{norm}}_n\mathrm{e}^{-\mathrm{i}\varphi(\nu_n; \delta\tau, \delta\tau_\text{iono})} - \enVert{F(\nu_n)}}^2}{\sigma_n^2}\nonumber\\
    &=& \sum^{N-1}_{n=0} \frac{1}{\sigma_n^2} \text{Im}\left[ \del{\mathcal{V}_\text{norm}}_n\mathrm{e}^{-\mathrm{i}\varphi(\nu_n; \delta\tau, \delta\tau_\text{iono})} \right]^2, \label{eq:likelihood-chi2}
\end{eqnarray}
where $\varphi(\nu_n;\delta\tau, \delta\tau_\text{iono})$ is a phase model independent of the amplitude, flux of the source and time $t$. Notice that the real part of Eq.~\eqref{eq:likelihood-chi2} can be marginalized without losing information, since $\text{Re}\sbr{\enVert{F}} = \enVert{F}$ and all astrometric information is only contained in the phase of the visibility, $\text{Arg}\sbr{\boldsymbol{\mathcal{V}}_\text{norm}}$.

In effect, the fringe phase model, $\text{Arg}\sbr{F}$, can be represented as:
\begin{eqnarray}
    \varphi(\nu;\delta\tau, \delta\tau_\text{iono}) &=& \varphi(\nu;\delta\tau) + \varphi(\nu;\delta\tau_\text{iono}) \nonumber \\
    &=& 2\pi\delta\tau\nu + 2\pi\delta\tau_\text{iono}\nu \nonumber \\
    &=& 2\pi\delta\tau\nu +\frac{2\pi}{\nu}{k_\text{DM}\delta\text{DM}},
    \label{eq:vis_fit}
\end{eqnarray}
where the site differential DM is given by

\begin{equation}
    \Delta\text{DM}=\text{DM}_A - \text{DM}_B\leq\SI{1e-5}{\parsec\per\centi\m\cubed},
\end{equation}
and $\delta\text{DM}$ the difference of differential DM from target and reference:

\begin{equation}
    \delta\text{DM} = \Delta\text{DM}^\text{t} - \Delta\text{DM}^\text{r}.
\end{equation}
The right term Eq.~\eqref{eq:vis_fit} comes from the differential dispersive delay phase, $2\pi\nu\tau_\text{dd}$, from Eq.~\eqref{eq:dispersive_time}.
Notice that the ionospheric contribution can be measured in DM units (\si{\parsec\per\centi\m\cubed}) or Total Electron Content\footnote{Total Electron Content and Dispersion Measure have equivalent units: \begin{equation*} \SI{1}{\TECu} \equiv \SI{1e16}{\electron\per\m\squared}\equiv \SI{3.24e-7}{\parsec\per\centi\m\cubed}.\end{equation*}} (TEC) units (\si{\TECu}), and it will depend on the zenith angle at each site (ionosphere column).
Similarly to delays, we are only interested in its difference rather than the true value at each site.

The residual delay $\delta\tau$, which does not consider $\delta\tau_\text{iono}$, can be expressed as:

\begin{align}
    \delta\tau & = \tau_\text{total}^\text{t} - \tau_\text{total}^\text{r} -\delta\tau_\text{iono} \nonumber\\
    & = \tilde{\tau}_\text{geo}^\text{t} + \tilde{\tau}^\text{t}_\text{clock} - \del{\tilde{\tau}_\text{geo}^\text{r} +\tilde{\tau}^\text{r}_\text{clock} }\nonumber \\
    & = \delta\tau_\text{geo} - \delta\tau_{\text{geo}_0}+\delta\tau_\text{clock} - \delta\tau_{\text{clock}_0},
\end{align}
with $\delta$ always expressing target (t; FRB) minus reference (r; calibrator).

The procedure presented above follows the same approach developed in \cite{2021AJ....161...81L} but is expanded to an additional degree of freedom for $\chi^2$, where the differential delay contribution from the ionosphere is non-negligible at this baseline $b$.

The solution to the presented method can be achieved by either a two dimensional grid search or by a gradient search (nonlinear least squares minimization), although the latter needs high precision in the initial guessed values (see results from Section \ref{sec:towards_localization} in Figure \ref{fig:grid_search_sd_0}). A good initial estimate is the FFT over the normalized (integrated) visibility,

\begin{equation}
    \enVert{\boldsymbol{\rho}^\text{sf}_{A,B}}=\enVert{\text{FFT}\sbr{\boldsymbol{\mathcal{V}}_\text{norm}}},
    \label{eq:cross-correlation-strength}  
\end{equation}
which returns the cross-correlation strength (lag-correlation in Eq.~\ref{eq:rhosf_visibility}) prior to residual corrections being applied. 
Further, if dispersive delays weren't present in the phase, $\boldsymbol{\varphi}\sbr{n}$, the maximum likelihood solution for fringe fitting would be the constant value obtained from the Fourier Transform.
Then the observable shift over the lag-axis (Eq.~\ref{eq:lag-correlation}) is the lag-correlation, a good initial value for $\delta\tau$ or equivalently the center of a grid search.

\subsection{Timing Performance}
\label{sec:timing_performance}

The CHIME clock system is composed of a GPS disciplined ovenized crystal oscillator \citep{tm-4d} of \SI{10}{\mega\hertz} tempo and absolute time from GPS signal (all coming from the same unit). The clock speeds up and slows down while retrieving the GPS signal and adjusting to it.
Measurements to test the reliability of the current CHIME clock and ARO maser were performed while observing the PSR B0531+21 compact source over the course of a day and over multiple days. Here we compare the computed time difference from the cross-correlation process (time difference between PSR B0531+21 pairs) and the CHIME clock stability. Correlations follow the same analysis explained in Section \ref{sec:wideband_fringe_fitting} (forming visibilities and fringe fitting).
The latter is an independent analysis done exclusively at the CHIME clock (and independent of the PSR B0531+21 observations), and it includes a study of the DRAO maser pHMaser 1008 \citep{t4scince} (properties in Table \ref{tab:DRAO_maser}) with respect to the CHIME clock phase and its raw data calibration in long timescales.
The method includes a pipeline to collect raw ADC samples directly from the maser and generate the GPS timestamps associated to them as well as to correct for long scale variations done by the imperfections in the \SI{10}{\mega\hertz} rate of the CHIME clock.

\begin{table}[t]
    \centering
    \caption{DRAO maser properties \citep{t4scince}.}
    \label{tab:DRAO_maser}
    \begin{tabular}{ll}
        \tableline
        Operation frequency & \SI{10}{\mega\hertz}\\
        Allan variance (at \SI{1}{\s} and \SI{1}{\hertz} bandwidth) & \num{5e-13}\\
        Temperature sensitivity & $<\SI{1e-14}{\per\degreeCelsius}$\\
        \tableline
        \end{tabular}
\end{table}

\subsubsection{Clock stability single day}
\label{sec:clock_stability_single_day}

\begin{figure}[t]
    \centering
    \includegraphics{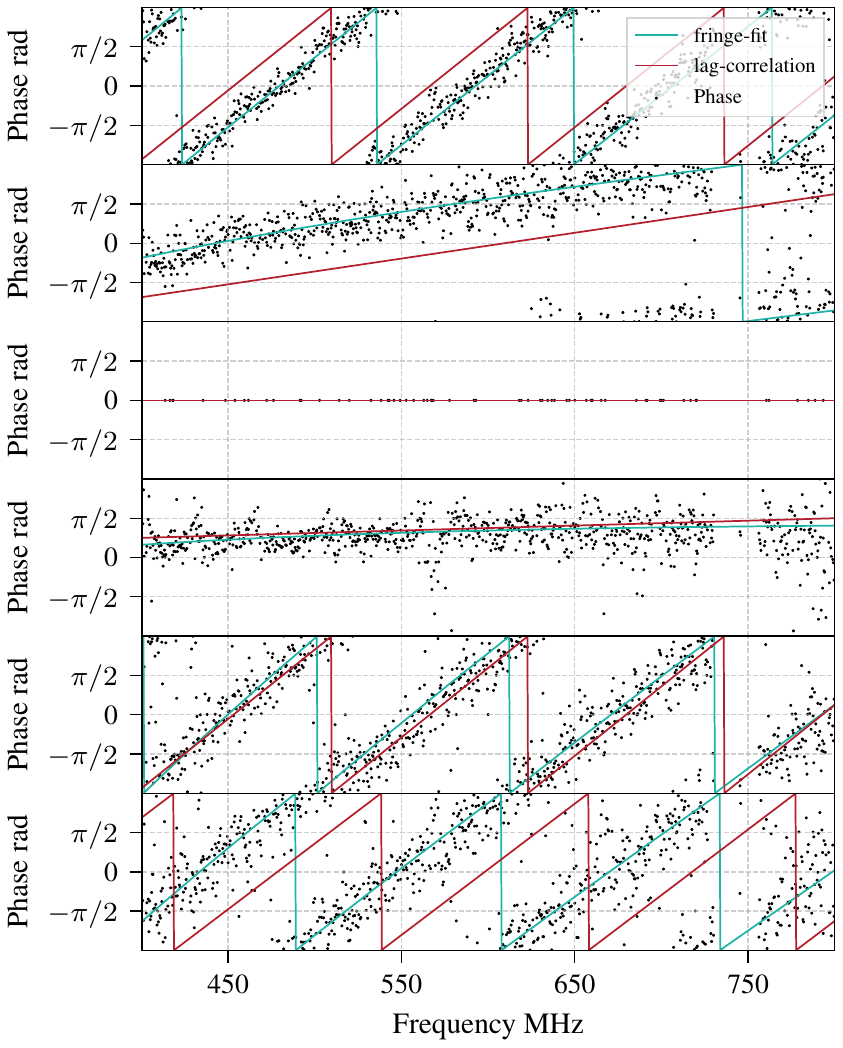}
    \caption{Phase of normalized (integrated) visibilities, $\boldsymbol{\varphi}\sbr{n}=\text{Arg}\sbr{\boldsymbol{\mathcal{V}}_\text{norm}}$, in a single day of observations for six different pulses from PSR B0531+21. Third panel has zero phase since the normalization is its own visibility. Wrapping lines are the phase of the models from fringe fitting and lag-correlation. Fringe fitting has an improved result since it takes into account the ionospheric contribution.}
    \label{fig:phases}
\end{figure}

During May \nth{9} 2020, six pulses of PSR B0531+21 were observed and baseband data collected at both sites with the CHIME VLBI tracking beam (defined in Section \ref{sec:triggering_from_chime}) and the ARO 10-m telescope, with each pulse separated by minutes. The cross-correlation follows the fringe fitting procedure explained in Section \ref{sec:wideband_fringe_fitting}, but instead of a pair of visibilities from the FRB and calibrator, the six pairs of pulses are referenced to the same pulsar, a single pulse (and the strongest) visibility of the same day\footnote{For the analysis in Sections \ref{sec:clock_stability_single_day} and \ref{sec:clock_stability_multiple_day},
only a single cross-polarization pair, $\mathbf{V}_{Y_A}$ and $\mathbf{V}_{X_B}$, are used to compute visibilities.}.
The PSR B0531+21 (reference) localization was used from the ATNF pulsar data set\footnote{\url{http://www.atnf.csiro.au/research/pulsar/psrcat}} \citep{2005AJ....129.1993M},
$\upalpha = \SI{5}{\hour}\SI{34}{\minute}\SI{31.973}{\s}$, and $\updelta=\SI{+22}{\degree}\SI{00}{\arcminute}\SI{52.06}{\arcsecond}$ (J2000.0).
Figure \ref{fig:phases} shows the wrapping phases of the normalized (integrated) visibilities:

\begin{equation}
    \boldsymbol{\mathcal{V}}_\text{norm} = \frac{\left<\boldsymbol{\mathcal{V}}_{Y_AX_B}^\text{t}\right>_{t_\text{w}}}{\left<\boldsymbol{\mathcal{V}}_{Y_AX_B}^\text{r}\right>_{t_\text{w}}} = \enVert{\boldsymbol{\mathcal{V}}_\text{norm}}\mathrm{e}^{\mathrm{i}\boldsymbol{\varphi}}, \label{eq:vis_ref_crab}
\end{equation}
with $\boldsymbol{\varphi}\sbr{n}=\text{Arg}\sbr{\boldsymbol{\mathcal{V}}_\text{norm}}$ a vector of $N$ channels. The best model fit can be achieved with Eq.~\eqref{eq:vis_fit}. The quantity in Eq.~\eqref{eq:vis_ref_crab} is computed six times for each pair of target pulses $\boldsymbol{\mathcal{V}}_{Y_AX_B}^\text{t}$. The matrix of complex numbers $\boldsymbol{\mathcal{V}}^\text{r}_{Y_AX_B}$ represents the strongest visibility (selected by measuring the amplitude of its Fast Fourier Transform, lag-correlation Section \ref{sec:correlation_with_baseband_data}), and for this observation it is the third pulse observed on May \nth{9}, 2020. Figure \ref{fig:phases} also contains the best fit from fringe fitting and lag-correlation phase models (colored lines).

Since pulses are separated only by a few minutes, the visibility normalization will remove most of the unknown delays (Eq.~\ref{eq:delays}), leaving only a small fraction of $\delta\tau$, and $\delta\tau_\text{iono}$ residuals. It's important to note that the ionospheric delay will depend on the zenith angle at each location \citep[see][Chapter~14]{2017isra.book.....T}, $\tau_\text{iono}=\tau_\text{iono}(z)$, and hence $\delta\tau_\text{iono}\neq\SI{0}{\nano\s}$. This implies that even if ionospheric turbulences are stable over an hour, there will still be a nonzero ionospheric delay since the differential zenith angle is changing over time.

\begin{figure}[t]
    \centering
    \includegraphics{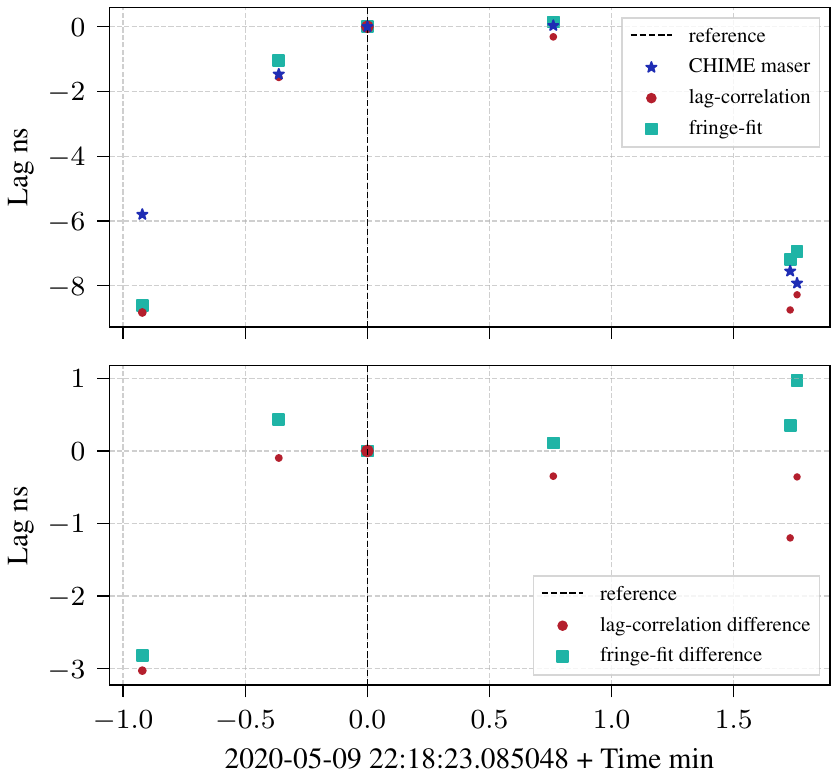}
    \caption{Clock stability between CHIME clock (GPS disciplined crystal oscillator) and ARO 10-m telescope over a single day with 6 pairs of correlated PSR B0531+21 pulses (separated over minutes scale). Circles are the lag-correlation between pulses (simple Fast Fourier Transform over the normalized integrated visibility), and the relative size represents the strength of the fringe amplitude. Stars are the CHIME clock and DRAO maser variations (independent measurement). Squares are the result from the fringe fitting process.
    All datasets are referenced to the third point from the left (dashed vertical line). The upper panel shows the comparison of the three data sets, and the lower panel shows the difference between lag-correlation and CHIME maser, and fringe fitting and CHIME maser.}
    \label{fig:clock_jittering_sd}
\end{figure}

Figure \ref{fig:clock_jittering_sd} shows the lag fringe fitting and lag-correlation (using Eq.~\ref{eq:cross-correlation-strength}), found in each of the phases of Figure \ref{fig:phases}. Figure \ref{fig:clock_jittering_sd} also shows the CHIME clock variations with respect to the DRAO maser (stars) plotted on top which are referenced to the same day \citep{2021RNAAS...5..216C}. The largest difference between them is roughly \SI{3}{\nano\s}. The bottom panel in Figure \ref{fig:clock_jittering_sd} shows the difference between the first order delay with lag-correlation and with fringe fitting. There exists a better estimation of the delay in the latter, since the $\delta\text{DM}$ has also been fitted.

In addition, the fringe fitting method (described in Section \ref{sec:wideband_fringe_fitting}) was applied to each normalized (integrated) visibility, and a Gaussian error distribution was assumed. To find the set of two parameters $(\delta\tau, \delta\text{DM})$ we calculated $\chi^2$ over a grid in expected DM and delay values giving lag-correlation as the search start point. The search was conducted over $\pm\SI{1e-6}{\parsec\per\centi\m\cubed}$ (centered at zero DM) and $\pm\SI{2.5}{\nano\s}$ centered at the peak of the lag-correlation.
The minimum over the grid is selected with a confidence region of \SI{68}{\percent}, $\chi^2 = \chi^2_\text{min} + \Delta\chi^2$, with $\Delta\chi^2=2.30$ for a two parameter model \citep{2012psa..book.....W}.

\subsubsection{Clock stability over multiple days}
\label{sec:clock_stability_multiple_day}

\begin{figure}[t]
    \centering
    \includegraphics{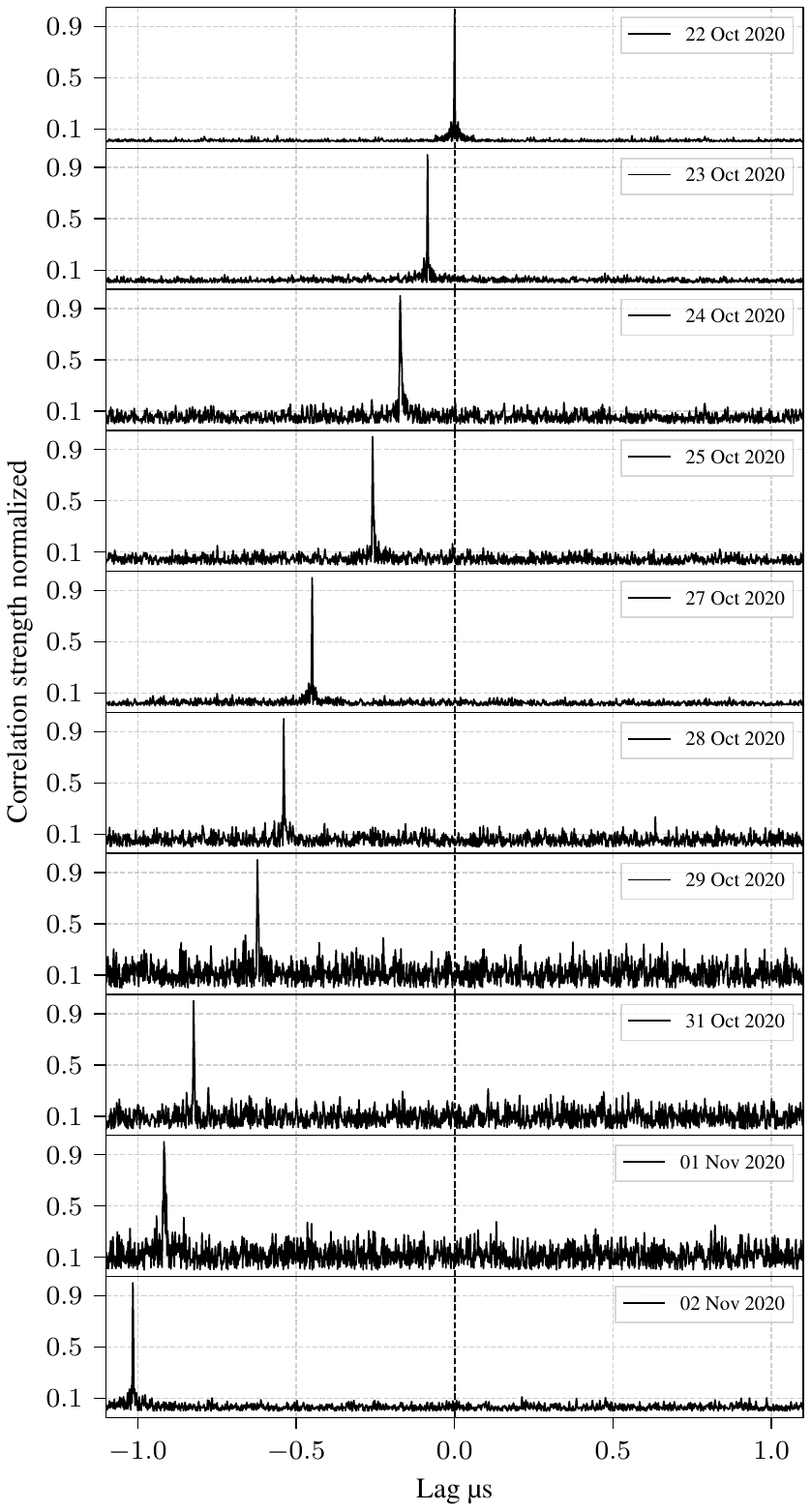}
    \caption{Fast Fourier Transform of normalized (integrated) visibilities $\enVert{\text{FFT}\sbr{\boldsymbol{\mathcal{V}}_\text{norm}}}$ {or equivalently the cross-correlation strength}, over several days for PSR B0531+21 pulses. The top panel is the reference pulse and the dashed line is the zero lag position. The lag-axis, $\tau_u$ elements, is of a frame size with limits $\pm\SI{1.28}{\micro\s}$. The observed drift in the cross-correlation peak is analyzed in Figure \ref{fig:clock_jittering_md}.}
    \label{fig:fft_lag}
\end{figure}

Using the same techniques as Section \ref{sec:clock_stability_single_day}, we now examine clock stability over multiple days using PSR B0531+21 observations of one pulse per day. Observations were performed with the CHIME/FRB baseband backend (defined in Section \ref{sec:triggering_from_chime}) and the ARO 10-m telescope dish. All pulses are referenced, $\boldsymbol{\mathcal{V}}^\text{r}_{Y_AX_B}$, to the first day of observation, following the fringe fitting algorithm and forming visibilities. Figure \ref{fig:fft_lag} shows the Fast Fourier Transform norm of the normalized (integrated) set of visibilities, where the strongest delay component (lag-correlation) is clearly visible as a peak. The phase graph of those visibilities is omitted because it will not add any helpful visualization, due to the fact that a lag on the order of \SI{1}{\micro\s} wraps the phase over \num{400} times.

\begin{figure}[t]
    \centering
    \includegraphics{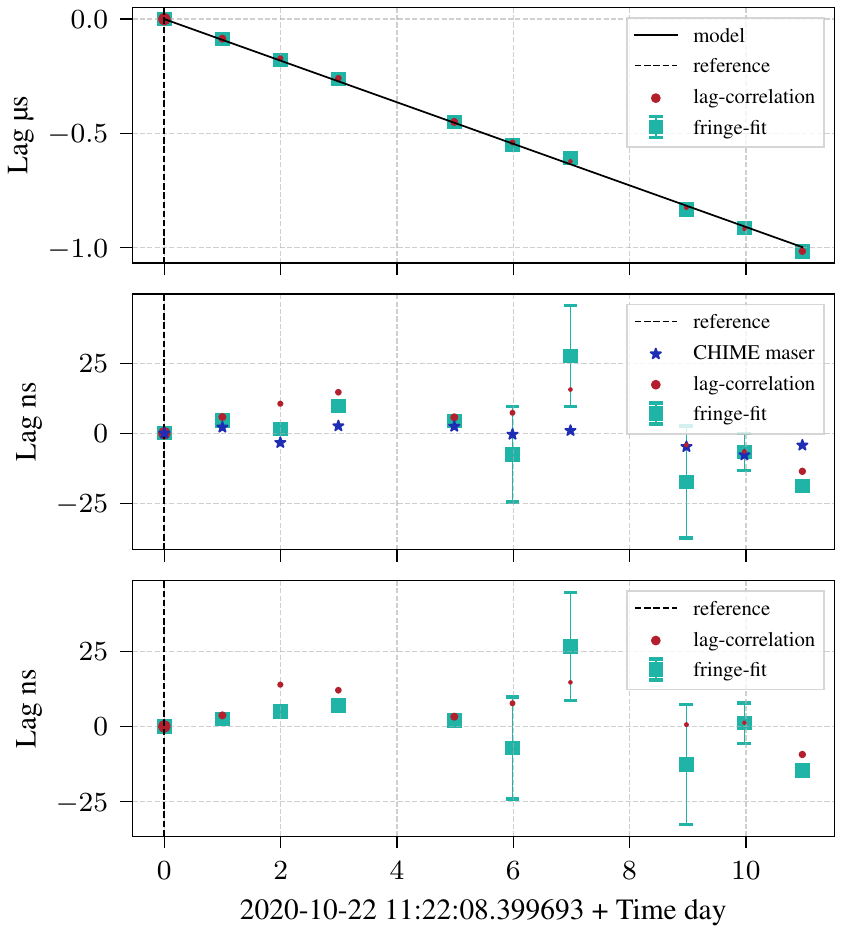}
    \caption{Clock stability between CHIME and ARO 10-m telescope over multiple days. The top panel shows the delays referenced to the first point at $(0, 0)$ (peak respect to the center in Figure \ref{fig:fft_lag}), and the sloped solid line is the best fit for the time drift. The middle panel shows the DRAO maser with respect to CHIME clock (stars) compared to the residual from the top panel (lag-correlation and fringe fitting). The marker sizes represent the signal-to-noise ratio strength in the lag-correlation with respect to the reference visibility $(0, 0)$ point (dashed vertical line). The last observations have a lower signal-to-noise ratio and the fringe fitting algorithm becomes less constrained in finding DM and lag. The bottom panel is the difference between the CHIME GPS crystal oscillator respect to DRAO maser (independent from correlations) and the two methods.}
    \label{fig:clock_jittering_md}
\end{figure}

The same (lag-correlation) delays from Figure \ref{fig:fft_lag} are then plotted in Figure \ref{fig:clock_jittering_md}.
In the top panel, it can be seen that for every day of observation there is a continuous time drift of \SI{\sim0.1}{\micro\s\per\day} (model sloped solid line), the source of which is mainly due to the combined rate between the CHIME GPS crystal oscillator clock and ARO maser relative frequency offset. The slope has been adjusted to lag-correlation (circles) and fringe-fit (squares) independently.
Both clocks should be \SI{10}{\mega\hertz}, but it is not precise over the timescale of days, hence they drift apart.
The middle panel shows the difference between the best fitted model (sloped solid line top panel) of the lag-correlation trend (over several days) and the data points. The same procedure for fringe fitting data points was used (where errors in the best fitted line are included). The DRAO maser independent measurements with respect to the CHIME clock are also plotted (stars).
Finally, in the bottom panel, the difference between the DRAO maser and the two methods is computed. By comparing stars and the other two set of points (circles and squares), we see that there is a clear improvement in fringe-fit for earlier days, and only a few corrections need to be applied to the phase of the normalized (integrated) visibilities.
The data points from the \nth{28}, \nth{29}, \nth{31} October, and \nth{1} November 2020 are less well constrained due to higher noise in the observations (as seen in the Figure \ref{fig:fft_lag} noise floor) and expected clock inaccuracies over more than five days, hence a clear value in $\chi^2$ is hard to achieve (higher degeneracy in the $\chi^2$ surface).

\begin{figure}[t]
    \centering
    \includegraphics{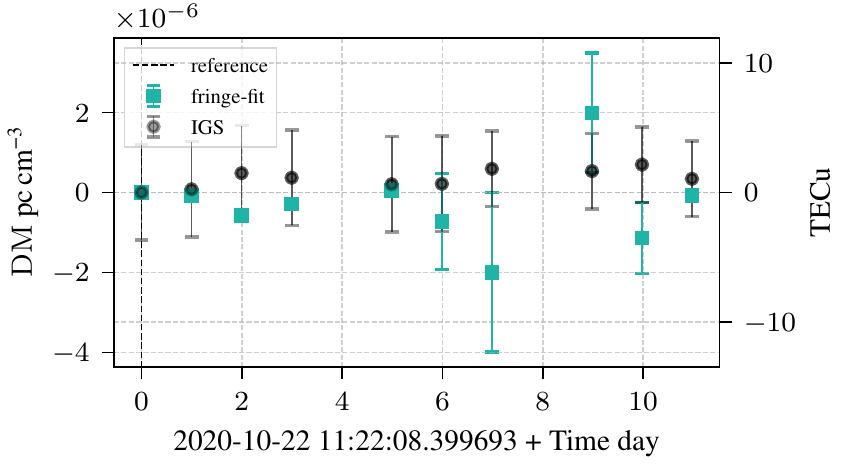}
    \caption{Ionospheric values found from fringe fitting (squares) compared to those same days in the IGS database \citep{NOLL20101421}. The dispersion measure values are the $\delta\text{DM}$ over the referenced day (first day vertical dashed line, \nth{22} October 2020), see Eq.~\eqref{eq:deltaDM_md}.
    Error bars in fringe fitting values are larger on the low signal-to-noise ratio cross-correlation cases i.e., \nth{28}, \nth{29}, \nth{31} October and \nth{1} November 2020. The TEC difference between sites was computed by using the slant TEC (sTEC). The fringe-fit $\delta\text{DM}$ agrees within errors compared to IGS, even after multiple days away from the referenced pulse.}
    \label{fig:ionosphere_md}
\end{figure}

The other $\chi^2$ degree of freedom is the $\delta\text{DM}$. This is not only the difference between sites, but since normalization took place (Eq.~\ref{eq:vis_ref_crab}), the computed value from minimizing $\chi^2$ is a differential $\Delta\text{DM}$:

\begin{equation}
    \delta\text{DM}^j=\Delta\text{DM}^j-\Delta\text{DM}^\text{r},\quad \text{with}\quad j=1,\dotso9,
    \label{eq:deltaDM_md}
\end{equation}
where $\Delta\text{DM}^\text{r}$ is the crossed ionosphere from the reference pulse and $j$ represents all other individual pulses' $\Delta\text{DM}$ (targets). We can then compare the results from the fringe fitting routine with the International Global Navigation Satellite System (GNSS) Service (IGS) \citep{NOLL20101421} which is able to compute TEC values to a precision of \SI{1e-6}{\parsec\per\centi\m\cubed} (or \SI{2.3}{\TECu}). The results are shown in Figure \ref{fig:ionosphere_md}.
TEC values interpolated with the ionospheric model take into consideration the zenith angle; this is referred to as slant TEC (sTEC), i.e.

\begin{equation}
    \Delta\text{TEC} = \text{sTEC}_A(z) - \text{sTEC}_B(z).
\end{equation}
The IGS published maps have a fiducial ionosphere height of \SI{450}{\kilo\m}, and we assume a thin layer approximation in translating the published vertical TEC maps to slant TEC, which estimates the TEC along the line of sight for each Earth location to the source.

\subsection{Towards Localization}
\label{sec:towards_localization}

For CHIME and the ARO 10-m telescope, we have shown that the combined clock systems will not drift apart more than \SIrange{10}{100}{\nano\s} over \SI{24}{\hour}. In addition, in the case where no calibrators are available over several days, the expected drift can be accounted for and removed. Nonetheless, limitations of the current coherent delay-referencing method (discussed in Section \ref{sec:discussion}), will not provide a full VLBI diffraction limited resolution.

Of utmost importance is the unknown contribution from the ionosphere, which depends on the Earth location and zenith angle of the observation. More generally, the ionosphere density along the lines of sight to the calibrator may be different from that along the lines of sight to the FRB.

\begin{figure*}
    \centering
    \includegraphics{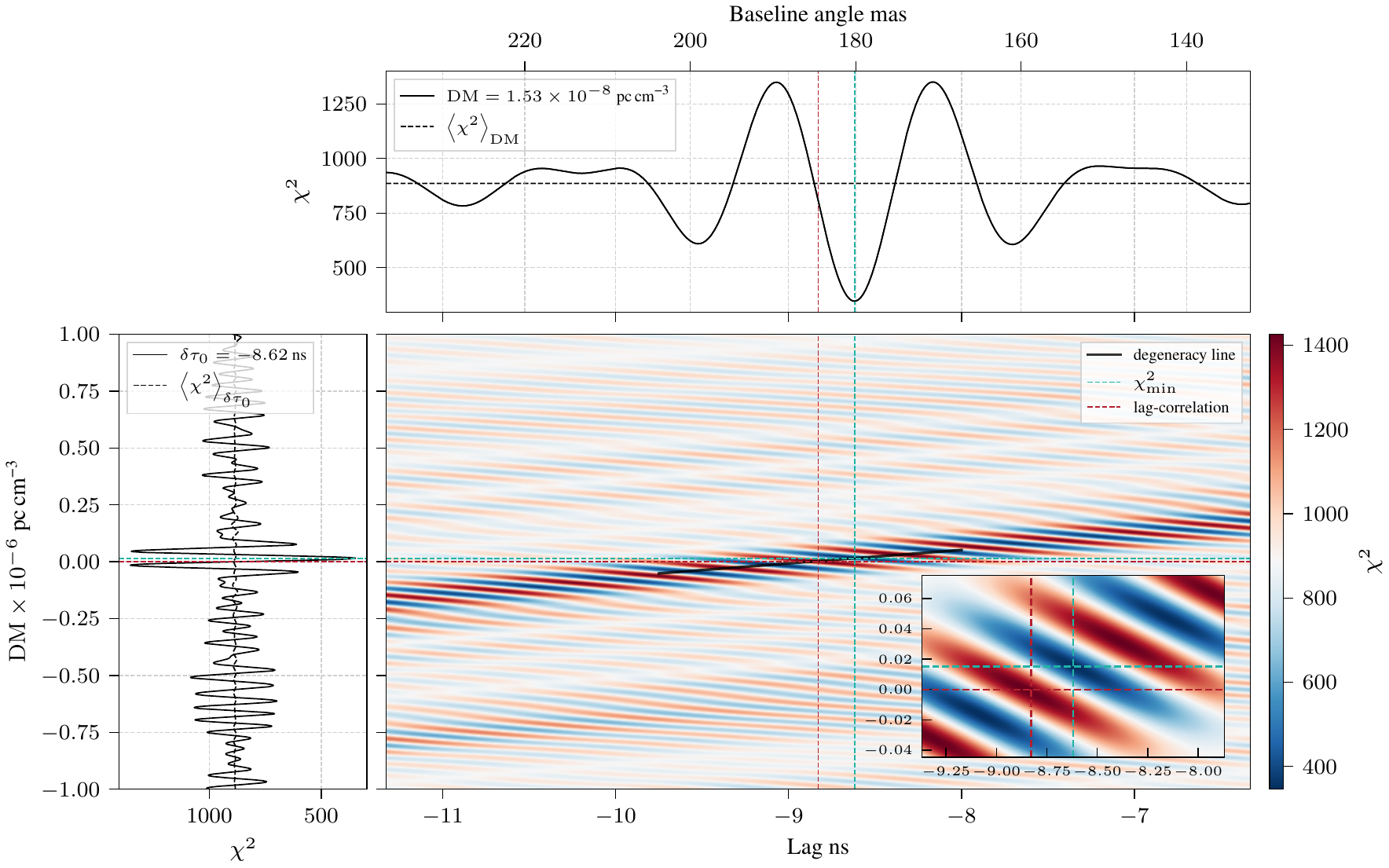}
    \caption{Grid search over lag and $\delta\text{DM}$ space for $\chi^2$. The solution corresponds to the first burst on the May \nth{9} 2020 observations (see Figure \ref{fig:clock_jittering_sd}), which is the largest deviated point from its reference (first data point from left to right). The top panel shows the slice of the optimum DM and the left panel the slice of the optimum lag ($\delta\tau$). The top panel top axis shows the differential baseline angle ($\delta\theta$) and it has directly translated from the obtained lag (see Eq.~\ref{eq:baseline_angle_delta}).
    The grid search is centered at the lag-correlation result (in this case \SI{-8.8}{\nano\s}) and zero DM, over $\pm\SI{1e-6}{\parsec\per\centi\m\cubed}$ and $\pm\SI{2.5}{\nano\s}$ windows. The bottom right zoom-in section is centered at the minimum $\chi^2$ value with axes units same as the major axes. Dashed lines cyan and red show the location of the $\chi^2_\text{min}$ and the lag-correlation. Figure \ref{fig:deg_line_sd_0} the degeneracy line or slice with most degeneracies is plotted with their confidence levels, and in Figure \ref{fig:grid_search_corner_sd_0} shows the same lobe in a corner plot using the obtained $\chi^2$ weights (likelihood).}
    \label{fig:grid_search_sd_0}
\end{figure*}

\begin{figure}[t]
    \centering
    \includegraphics{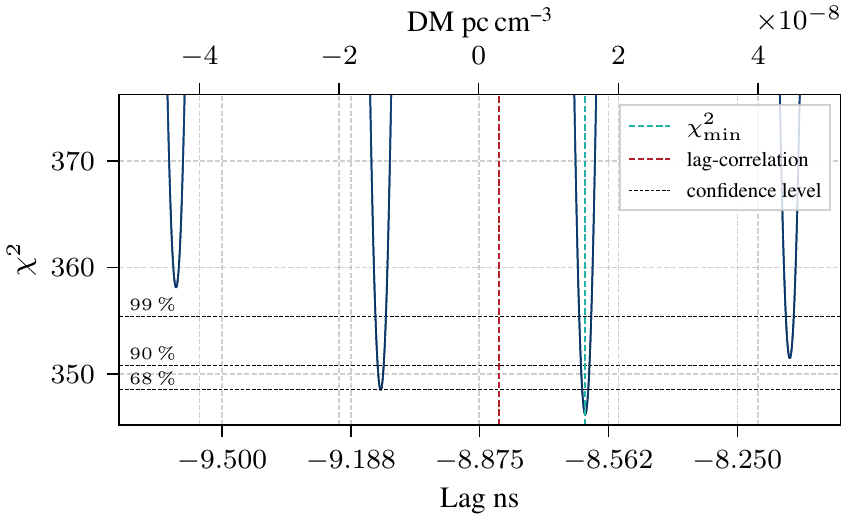}
    \caption{Degeneracy line, one dimensional slice from the $\chi^2$ grid in Figure \ref{fig:grid_search_sd_0}. The slice shows a zoom-in section of the grid search along the most degenerated lobes. Horizontal dashed lines are the confidence level from the $\chi^2_\text{min}$ value, \SIlist{68;90;99}{\percent}. Vertical dashed lines are the lag-correlation and $\chi^2$ minimum. The plot corresponds to the most deviated point in a single day observation (Section \ref{sec:clock_stability_single_day}), and the one that has most degeneracies.}
    \label{fig:deg_line_sd_0}
\end{figure}

From the analysis in Sections \ref{sec:clock_stability_single_day} and \ref{sec:clock_stability_multiple_day}, the fringe fitting finds an estimated ionosphere contribution provided that our observations have a strong signal-to-noise ratio. An example of a single point is presented in Figure \ref{fig:grid_search_sd_0} and \ref{fig:deg_line_sd_0}, which shows the $\chi^2$ surface over $\delta\text{DM}$ and lag space (grid search), and a zoom-in or slice section of $\chi^2$. The correlated pair corresponds to the first burst in Figure \ref{fig:clock_jittering_sd} observed on May \nth{9} 2020, which has the largest scatter among all points (mainly due to the error in the CHIME GPS crystal oscillator) in a single day observation.

The minimum $\chi^2$ in Figure \ref{fig:grid_search_sd_0} is located at the center with cyan dashed lines. It is not surprising that the surface is highly irregular with neighboring values that could reach an equal level of degeneracy (as seen in the nearby lobes from Figure \ref{fig:deg_line_sd_0}), but a high enough signal-to-noise ratio and a good prior in lag and DM can solve this problem (no prior information was used in fringe fitting Sections \ref{sec:clock_stability_single_day} and \ref{sec:clock_stability_multiple_day}).

To coherently find the delay-referenced localization of a target pulse with respect to the reference pulse, we calculate it as a function of the residual baseline angle $\delta\theta$ (defined in Figure \ref{fig:VLBI_basics} and Eq.~\ref{eq:baseline_angle}), i.e.,

\begin{gather}
    \varphi = 2\pi\delta\tau\nu \approx 2\pi\delta\tau_\text{geo}\nu = 2\pi b\frac{\nu}{c}\del{\cos\theta_\text{t} - \cos\theta_\text{r}},\nonumber\\
    \frac{c\delta\tau_\text{geo}}{b} = \cos\theta_\text{t} - \cos\theta_\text{r}= -\delta\theta\sin\theta_\text{r},\nonumber\\
    \delta\theta = -\frac{c\delta\tau_\text{geo}}{b\sin\theta_\text{r}}
    \label{eq:baseline_angle_delta}
\end{gather}
where $\delta\theta(\upalpha,\updelta) = \theta_\text{t}-\theta_\text{r}$ (difference from target pulse with respect to reference pulse on sky) is a function of RA and DEC \citep[see][Chapter~12]{2017isra.book.....T}, but higher restrictions will be given to RA due to the East-West baseline. The angle of the normalized (integrated) visibility $\varphi\approx\varphi_\text{geo}$ is the same as defined in Eq.~\eqref{eq:phase_ref} but now the small angle approximation has been used for $\delta\theta$.
The reference angle, $\theta_\text{r}$, is the known position of the calibrator and (as with the geometric delay) it will depend on: Earth locations, observation time ($t_A$), and RA and DEC of the reference source.
Then we simply use Eq.~\eqref{eq:baseline_angle} and compute $\tau_\text{geo} = \tau_{\text{geo}_0}$ (\texttt{difxcacl11}) since there will be no residual using the assumed reference pulse. For the single day observations (Section \ref{sec:clock_stability_single_day}) this value corresponds to $\theta_\text{r}=\SI{1.29}{\radian}$.
Eq.~\eqref{eq:baseline_angle_delta} will only be true when the residual of the clock delay $\delta\tau_\text{clock}$ is very small, viz., $\delta\tau\approx\delta\tau_\text{geo}$ (same as in Eq.~\ref{eq:baseline_angle}, Section \ref{sec:vlbi_between_the_aro_10-m_and_chime}).

\begin{figure}[t]
    \centering
    \includegraphics{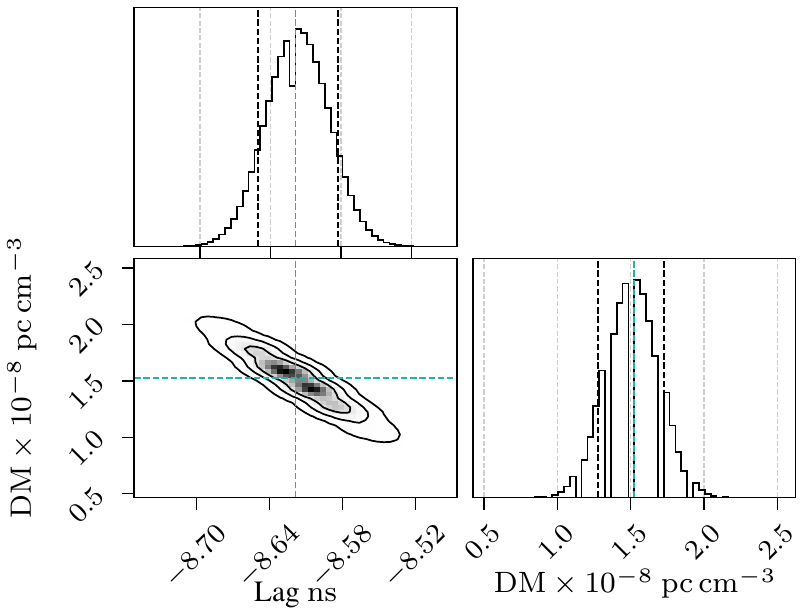}
    \caption{Example of a well-constrained minimum $\chi^2$ lobe from Figure \ref{fig:grid_search_sd_0}. The centered dashed line is the minimum $\chi^2$ (or equivalently maximum $\mathcal{L}$). The quantiles (\SI{10}{\percent} and \SI{90}{\percent}) show a rough estimate required for a good cross-correlation in $\delta\text{DM}$ space of \SI{1e-8}{\parsec\per\centi\m\cubed}. Contour lines correspond to \SIlist{68;90;99}{\percent} confidence. The lobe is also shown in Figure \ref{fig:deg_line_sd_0}, where it is compared to the closer and most degenerated lobes in the degeneracy line from the entire $\chi^2$ grid (Figure \ref{fig:grid_search_sd_0}).
    The example corresponds to the first burst (left to right) from Figure \ref{fig:clock_jittering_sd}.}
    \label{fig:grid_search_corner_sd_0}
\end{figure}

\begin{figure}[t]
    \centering
    \includegraphics{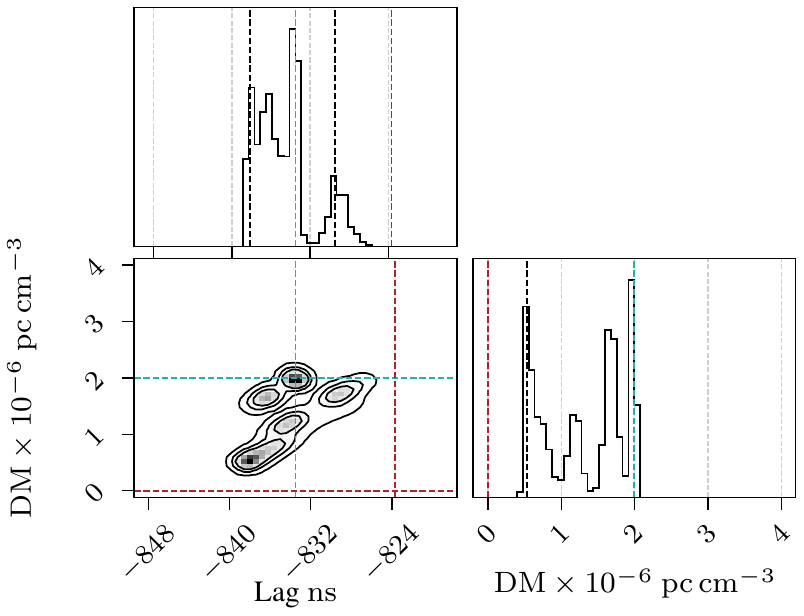}
    \caption{Differential DM and lag selection of the minimum $\chi^2$ lobe. Large uncertainties are due to the poor signal-to-noise ratio of pulses and maser clock drift over several days. The corner plot correspond to \nth{31} October 2020 observation. Multiple lobes are formed in the case were a true optimum cannot be determined and non-gaussian uncertainties are high. Black dashed lines correspond to the \SI{10}{\percent} and \SI{90}{\percent} quantiles, the center line is the obtained minimum $\chi^2$, and off-center dashed lines are the grid search center given by the lag-correlation. Contour lines correspond to \SIlist{68;90;99}{\percent} confidence.}
    \label{fig:grid_search_corner_md_7}
\end{figure}

The top panel top axis from Figure \ref{fig:grid_search_sd_0} shows the baseline angle, $\delta\theta$, with respect to the delay $\delta\tau$ (centered panel bottom axis), and for this case the localization error is \SI{180}{\mas}. Fringe fitting errors can also be estimated: Figure \ref{fig:grid_search_corner_sd_0} shows the main lobe from Figure \ref{fig:grid_search_sd_0} with a sub-nanosecond error, and a \SI{1e-8}{\parsec\per\centi\m\cubed} DM. The \SI{180}{\mas} error in localization represents the largest scatter (as seen in the first point of Figure \ref{fig:clock_jittering_sd}).
In contrast to single day observations, the worst case over multiple days (\nth{31} October) is in Figure \ref{fig:grid_search_corner_md_7}, which shows the largest scatter from all points mainly due to the low signal-to-noise ratio in the lag-correlation and calibration after several days. In particular, days \nth{28}, \nth{29}, \nth{31} October and \nth{1} November 2020 do not have a clear Gaussian lobe and their error bars were estimated from the \SI{99}{\percent} ($\chi^2_\text{min} + 9.21$) confidence region (Figures \ref{fig:clock_jittering_md} and \ref{fig:ionosphere_md}).

In general, errors in fringe fitting over a single day and even over a few consecutive days are within $\SI{\pm10}{\nano\s}$ from the expected CHIME clock drift (with the exception of low signal-to-noise ratio over multiple days running clock). Beyond that, the combination of clocks is not reliable. Careful measurements from clocks at each station can improve the performance beyond a \SI{50}{\mas} localization by adjusting the delay from the known CHIME GPS crystal oscillator jitter.

\section{Early Science Results}
\label{sec:early_science_results}

As mentioned earlier, CHIME/FRB statistics predict that we should observe \SIrange{\sim4}{7}{\FRBs\per\month} in the ARO 10-m telescope and CHIME shared FoV. However, this does not take into account the up time from CHIME and the ARO 10-m telescope combined, plus the estimated altitude, azimuth, and Earth location of the 10-m telescope, i.e., the total portion covering the CHIME FoV may not be complete. Nevertheless, FRBs have been recorded in simultaneous dumps at CHIME and ARO 10-m telescope but unfortunately their signal-to-noise ratio ($\text{SNR}_\text{CHIME}<25$) on the ARO site was insufficient for both visualization in data and cross-correlations.
One of the  most prominent observation produced thus far is the magnetar SGR 1935+2154 \citep{2020Natur.587...54C}, where the ARO 10-m telescope recorded baseband data but CHIME unfortunately did not.

\begin{figure}[t]
    \centering
    \includegraphics{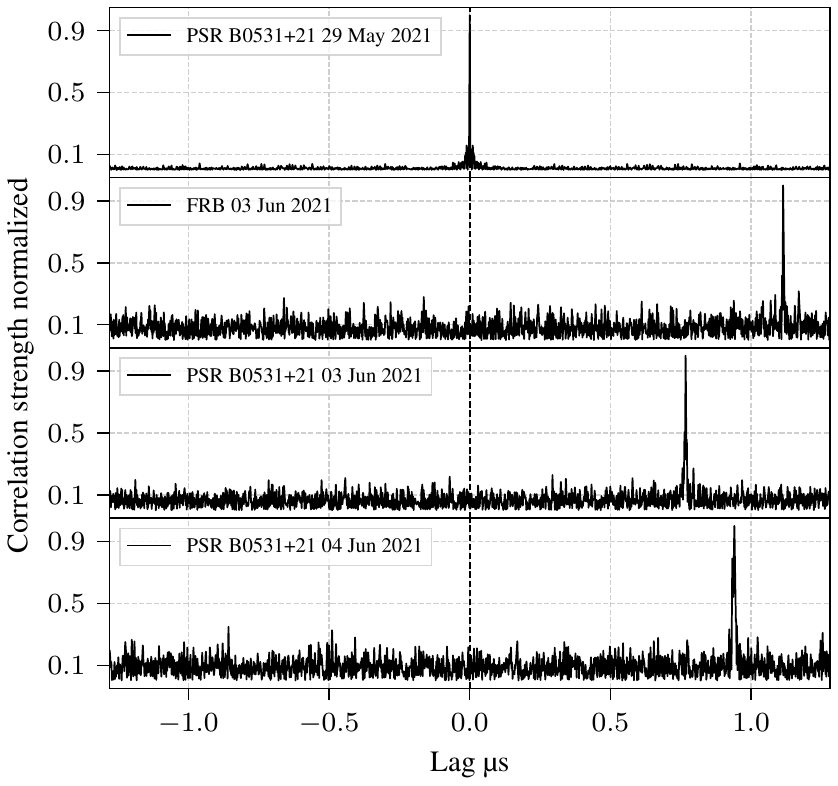}
    \caption{First FRB 20210603A detected and correlated at CHIME and ARO 10-m telescope in baseband data. The plot is similar as in Figure \ref{fig:fft_lag}, where we computed the cross-correlation strength (norm of the sub-frame cross-correlation function Eqs.~\ref{eq:rhosf_visibility} and \ref{eq:cross-correlation-strength}).
    The plot shows three pulses from PSR B0531+21 and the single FRB pulse. The PSR B0531+21 (referenced to 29 May 2021) has a similar trend to that seen in Figure \ref{fig:fft_lag}, but only over the pulsar pulses, since the FRB burst is expected to be at a different RA sky location. The normalized (integrated) visibility phase can be seen in Figure \ref{fig:phases_frb}.}
    \label{fig:fft_lag_frb}
\end{figure}

Lastly, observations from a the FRB 20210603A (not known to repeat), were recorded simultaneously at CHIME and ARO 10-m telescope on 2021 June \nth{3} and initial results show a strong cross-correlation after delay-referencing to a calibrator (PSR B0531+21 pulse). Figure \ref{fig:fft_lag_frb} shows the cross-correlation strength (Eq.~\ref{eq:cross-correlation-strength}), where the reference pulse is in the top panel. The known drift between the CHIME and ARO 10-m clocks can again be seen in the PSR B0531+21 pulses (similar to Figure \ref{fig:fft_lag}), but not in the FRB since its sky location is different.

The FRB burst, not visible at the ARO 10-m telescope, was triggered and successfully correlated hours after the event happened.
The FRB 20210603A had a high signal-to-noise ratio at CHIME of $\gtrsim100$ and $\text{DM}=\SI{500.162 +- 0.005}{\parsec\per\centi\m\cubed}$ (with a dispersion time $t_\text{dd}=\SI{9.728}{\s}$ over the band).

\begin{figure}[t]
    \centering
    \includegraphics{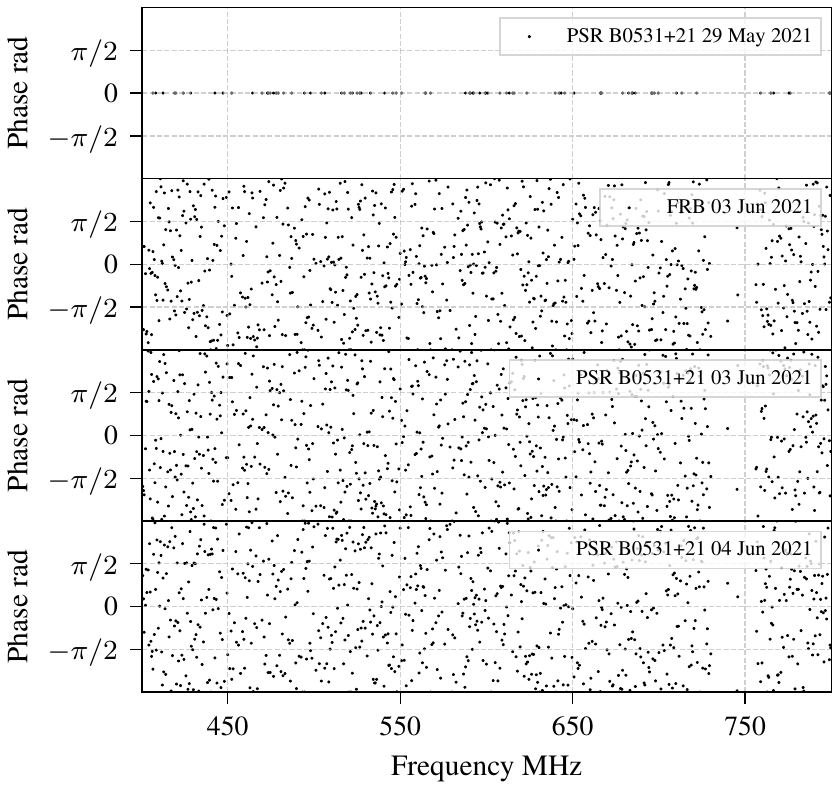}
    \caption{Normalized (integrated) visibility phase $\text{Arg}\sbr{\boldsymbol{\mathcal{V}}_\text{norm}}$ from PSR B0531+21 pulses and FRB 20210603A. Phases have been referenced to the PSR B0531+21 pulse on 2021 May 29, and there is no correction applied on them. Figure \ref{fig:phases_frb_fs} shows the same phases but corrected by the estimated lag $\tau_\text{lc}$ Eq.\ \eqref{eq:frb-fs}, a partial fringe stop.}
    \label{fig:phases_frb}
\end{figure}

The phase of the normalized (integrated) visibility $\text{Arg}\sbr{\boldsymbol{\mathcal{V}}_\text{norm}}$ can be seen in Figure \ref{fig:phases_frb}. No clear signal is visible since phases are wrapping fast (see Section \ref{sec:clock_stability_multiple_day}) due to the CHIME clock and ARO maser combination, and due to the difference in sky location of the FRB (only applicable in the second panel from top to bottom).

\begin{figure}[t]
    \centering
    \includegraphics{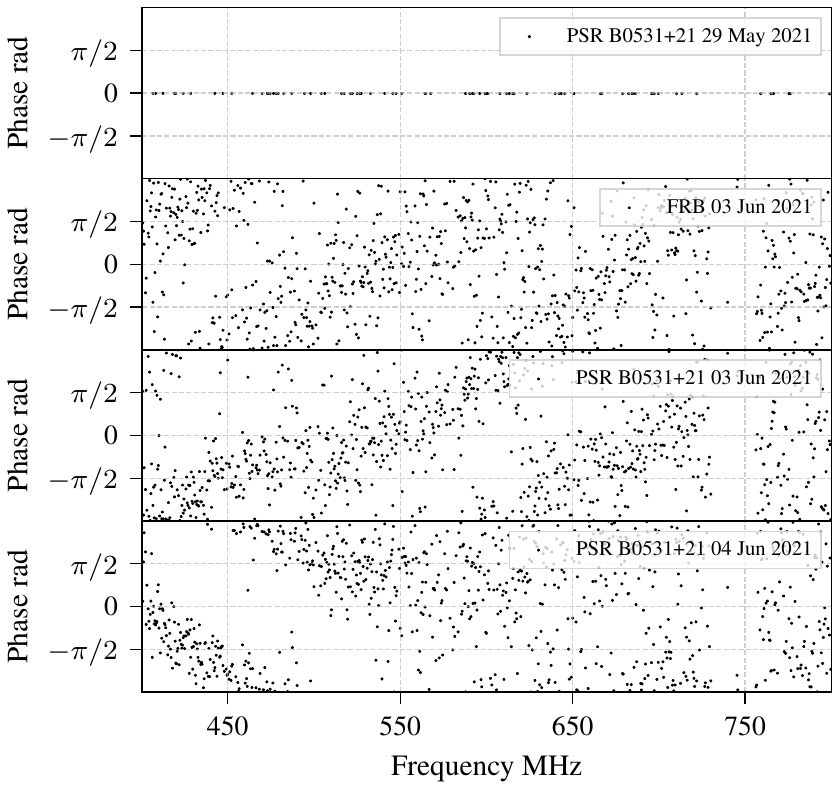}
    \caption{Normalized (integrated) visibility phase after phase correction (Eq.~\ref{eq:frb-fs}) from PSR B0531+21 pulses and FRB 20210603A. As in Figure \ref{fig:fft_lag_frb}, phases have been referenced to the pulse from 29 May 2021. In this early analysis only a phase correction over a constant delay (partial fringe stop) has been applied and the ionospheric correction has been set free (not corrected). The linear and nonlinear wrapping can be seen since the strong ionosphere has not been removed. The early study shows a clear detection of the FRB referenced to a calibrator (second panel top to bottom).}
    \label{fig:phases_frb_fs}
\end{figure}

By taking the estimated delay from the lag-correlation (Figure \ref{fig:fft_lag_frb}), we can partially fringe stop the phases in Figure \ref{fig:phases_frb}. This is:

\begin{equation}
    \del{\mathcal{V}_\text{norm}}_n \times \mathrm{e}^{-\mathrm{i}2\pi \nu_n \tau_\text{lc}}, \label{eq:frb-fs}
\end{equation}
with $\tau_\text{lc}$ the solution obtained from the sub-frame cross-correlation function, the lag-correlation (Eq.~\ref{eq:lag-correlation}).
There will be four $\tau_\text{lc}$; one for each panel.
Figure \ref{fig:phases_frb_fs} shows the angle of the normalized (integrated) visibility after being corrected by the delay found in Figure \ref{fig:fft_lag_frb}. This early study shows a clear detection of an FRB referenced to a calibrator as seen in Figure \ref{fig:phases_frb_fs} (second panel from top to bottom).

A more in depth analysis will be carried out in a separate study showing the proper localization results as well as the ionospheric and long dispersion corrections required at these frequencies.

\section{Discussion}
\label{sec:discussion}

The VLBI experiment from CHIME and the ARO 10-m telescope at low frequencies presents several challenges that have been partially addressed in the previous sections. In particular, we have addressed the clock stability: maser and GPS crystal oscillator, and the implementation of the fringe fitting algorithm (correlator model) with two degrees of freedom ($\delta\tau$ and $\delta$DM) and the localization uncertainty, which is mainly due to the ionospheric delay.
It follows the discussion of other important uncertainty contributions to the experiment: clock stability (during target and reference), ionosphere and troposphere delay, error in antenna positions (phase center interferometer), and Earth orientation parameters.

\subsection{Clock stability}

The CHIME GPS crystal oscillator is a GPS disciplined clock, which returns absolute timestamps but lacks the nanosecond precision required for VLBI. In contrast the DRAO maser fulfills the nanosecond precision but it does not support absolute GPS timestamps. A VLBI GPS disciplined maser is expensive and not feasible for the scale of the ARO 10-m telescope testbed. Nevertheless, our analysis has demonstrated that such deviations can be, in principle, corrected (e.g., by measuring pulsars, Figures \ref{fig:clock_jittering_sd} and \ref{fig:clock_jittering_md}); and measuring the CHIME GPS crystal oscillator with respect to the DRAO maser \citep{2021arXiv211000576M,2021RNAAS...5..216C}. Such measurements can measure uncertainties and correct clock variations to achieve a near \SI{\sim50}{\mas} for the future CHIME/FRB Outriggers project. 

The limiting factor for this calibration will be given by the FRB position and its proximity to the Galactic Plane (higher pulsar density) as well as the signal-to-noise ratio of both the calibrator and the source.

\subsection{Ionospheric contribution}

The ionospheric delay is a problem that has been partially solved in the analysis sections. This is due to the fact that for FRB detections, it is highly unlikely that the event will cross the same portion of the ionosphere as the calibrator. In reality, the two will likely be off by a few degrees in the sky, and the calibrator will be observed minutes before or after the event. In contrast, the analysis presented here relied on single pulses from the same pulsar as both target and reference, and the only difference in $\Delta\text{DM}^\text{t}$ and $\Delta\text{DM}^\text{r}$ are cases given by: the zenith angle over minutes (Section \ref{sec:clock_stability_single_day}) and day-to-day ionospheric changes (Section \ref{sec:clock_stability_multiple_day}). In the case of an FRB, we expect a higher degree of uncertainty since $\boldsymbol{\mathcal{V}}_\text{norm}$ will have a considerable fraction of residual ionospheric delay, $\delta\tau_\text{iono}$, in its phase after normalizing (Eq.~\ref{eq:vis_ref_crab}). Ionospheric models such as IGS \citep{NOLL20101421} do not have the required TEC precision that we need to successfully correlate at a desired angular precision of \SI{50}{\mas}, and $\delta\text{DM}\sim\SI{1e-8}{\parsec\per\centi\m\cubed}$ (Figure \ref{fig:grid_search_corner_sd_0}). However, high enough signal-to-noise ratio levels can partially remove this restriction and return a clear minimum in the $\chi^2$ domain.
An additional resource to improve the search is to add a prior probability over the ionospheric delay, using time-series GPS data to estimate the scale and type of ionosphere fluctuations (\SI{\sim10}{\TECu}). Adding this prior in the $\chi^2$ search can be expressed as:

\begin{equation}
    \chi^2_\text{eff} = \chi^2 + \chi^2_\text{prior}\del{\Delta\text{DM}^\text{t}, \Delta\text{DM}^\text{r}},
\end{equation}
where $\chi^2_\text{eff}$ is the effective $\chi^2$ weighted by an extra term, $\chi^2_\text{prior}$, which can be a function of the ionospheric delay or differential DM.

\subsection{VLBI delay and rate}
\label{sec:vlbi_delay_and_rate}
In practice the geometric phase correction, $\boldsymbol{\phi}_{\text{geo}_0}$ (Eq.~\ref{eq:baseband_geo_corr}), applied to the data (Section \ref{sec:wideband_fringe_fitting}, step 2) partially corrects for the delay rate. Since the location of PSR B0531+21 is well known, the delay rate $\frac{\partial\tau}{\partial t}$ vanishes for the analysis thus far. However, a significant delay rate can cause a drop in sensitivity because we correlate over the pulse width (which is on the order of a
millisecond) per frequency channel. This delay rate can be significant and is on the order of,

\begin{equation}
    \frac{\partial\phi}{\partial t} = 2\pi\nu\frac{\partial\tau}{\partial t} \Rightarrow \frac{\partial\tau}{\partial t}\approx \frac{\Delta\tau_\text{geo}}{\SI{1}{\milli\s}} = \SI{0.667}{\micro\s\per\s}.
\end{equation}
This estimate was computed while using a PSR B0531+21 pulse with the software \texttt{difxcalc11} using the same baseline $b$, i.e.,

\begin{equation}
   \Delta\tau_\text{geo} = \tau_{\text{geo}_0}(t_1 + \SI{1}{\milli\s}) - \tau_{\text{geo}_0}(t_1),
   \label{eq:delta_tau_geo}
\end{equation}

To solve this problem, the Global Fringe Fitting approach developed by \citet{1983AJ.....88..688S} for VLBI could be used. In this method, the phase model is explicitly dependent on time,

\begin{equation}
    \phi(t, \nu) = \phi(t_0, \nu_0) + \frac{\partial\phi}{\partial t}(t-t_0) + \frac{\partial\phi}{\partial \nu}(\nu-\nu_0),
\end{equation}
where the term $\frac{\partial\phi}{\partial t}$ is the fringe rate and $\frac{\partial\phi}{\partial \nu}$ the delay (or delay residual). The continuous phase distribution $\phi$ is analogous to $\boldsymbol{\phi}\sbr{n, k}$ of the visibility (Eq.~\ref{eq:vis_phi}). When the ionospheric contribution also cannot be neglected it may be necessary to use a model of the form

\begin{equation}
\phi(t, \nu) = \phi(t_0, \nu_0) + 2\pi\nu\tau + 2\pi\nu\frac{\partial\tau}{\partial t}t + \frac{2\pi}{\nu}k_\text{DM}\Delta\text{DM},
\end{equation}
to capture the delay rate within each channel.

\subsection{Earth locations and phase center of an interferometer}

Acquiring a good localization requires knowing our baselines to within a centimeter precision, which is not the case for the ARO 10-m telescope (the location of which is only known up to a \SI{\sim1}{\m} precision). Additionally, CHIME and future outrigger stations will need to account for the fact that the phase center location of CHIME at the time of beamforming may vary. These variations can result in the Earth location moving on the order of centimeters, which makes the geometric delay correction uncertain. Variations of this scale will have an impact within a frame (below \SI{2.56}{\micro\s}, but multiple nanoseconds), and will result in degeneracies in the FRB localization search. However, this problem can be solved by observing two or more calibrators daily (pulsars and expanding to VLBI calibrators for the CHIME/FRB Outriggers project), and instead of fringe fitting over a source, we use these observations to find the baseline positions. 

\subsection{Ionospheric Faraday rotation}

The ARO 10-m telescope has been used to study polarization properties \citep{2020Natur.587...54C}, but a linear polarization calibration (e.g., daily monitoring of a weakly polarized source) to solve for polarization leakage and other errors has not previously been carried out. This implies that a true transformation from the Cloverleaf feed linearly polarized to circular polarizations ($\mathbf{V}_{R_B}, \mathbf{V}_{L_B}$) is not obvious, and needs a complex gains correction.
For longer baselines it may be important to work in a circular basis since ionized plasma in the ionosphere could produce Faraday rotation (e.g., when the Sun's activity is high) and add a constant offset in right-left phase difference \citep{1995ASPC...82..289C}, where this effect has been noted at lower frequencies than \SI{400}{\mega\hertz}. Additionally, note that we don't have a good estimate of the North-South or East-West estimate of the linear polarizations orientation at the ARO 10-m telescope feed. This treatment has been neglected in the present analysis.

\subsection{Geometric delay and high DM}
\label{sec:geometric_delay_and_high_dm}

\begin{figure}[t]
    \centering
    \includegraphics{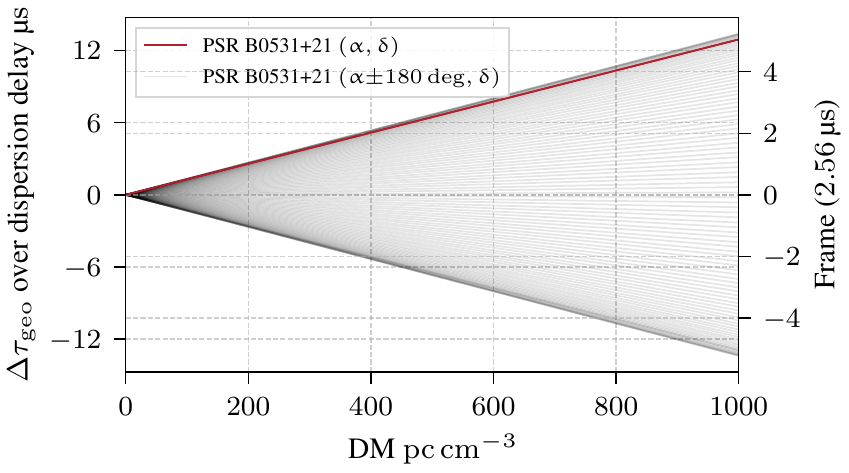}
    \caption{Geometric delay difference over the start (\SI{800}{\mega\hertz}) and stop (\SI{400}{\mega\hertz}) over a pulse dispersion, $\Delta\tau_\text{geo}$. The geometric delay was computed using PSR B0531+21 coordinates (CHIME and ARO 10-m telescope baseline) and varying the RA component $\SI{\pm180}{\deg}$.}
    \label{fig:geo_delay_frames}
\end{figure}

FRB events detected at CHIME and the ARO 10-m telescope will also have to contend with a larger dispersion which will cause large differences (compared to PSR B0531+21) in geometric delay at the top and bottom of the \SIrange[range-units=single, range-phrase=--]{400}{800}{\mega\hertz} band. As stated in the geometric correction method (Section \ref{sec:wideband_fringe_fitting}), the integer part of the geometric delay (floor part) in terms of frames $\left\lfloor\frac{\tau_{\text{geo}_0}}{\SI{2.56}{\micro\s}}\right\rfloor$ may not be constant along the pulse over each frequency channel, hence the data aligned between correlated polarizations will change slightly as well as the sub-frame correction over the phase in the pulse's coherent dedispersion. This effect becomes relevant at $\text{DM}\gtrsim\SI{200}{\parsec\per\centi\m\cubed}$ at comparable baselines. Figure \ref{fig:geo_delay_frames} shows the effect of the geometric delay over the dispersion delay. The quantity $\Delta\tau_\text{geo}$ is the difference of geometric delay (CHIME and ARO 10-m telescope baseline) at the start and end of the pulse dispersion (\SI{400}{\mega\hertz} bandwidth). The calculations were done on \texttt{difxcalc11} centered at PSR B0531+21 coordinates, similarly as in Eq.~\eqref{eq:delta_tau_geo}.
A special treatment of the band may be required: for example, separating the band in two or more sections with a $k_\text{shift}\sbr{n}$, forming visibilities, and then concatenating phases; but in general combining a coherent dedispersion with a time variable geometric delay is not a trivial problem to solve.

As explained in Section \ref{sec:wideband_fringe_fitting}, we are interested in the differential DM between sites, which is more than just the ionospheric contribution at sites $A$ and $B$. There will also be a Doppler effect given by the Earth's rotation respect to the Solar System Barycenter (SSB) which will impact the velocity of each telescope \citep{1970ApJ...162..707R, 2014ApJ...790...93P}. This results in $\text{DM}\to\widetilde{\text{DM}}$, a modified DM,

\begin{equation}
    \widetilde{\text{DM}}(v) = \text{DM} \frac{1}{1 + \frac{v}{c}},\label{eq:dm_vel}
\end{equation}
with $c$ the speed of light, and $v$ the velocity of one site. The Eq.~\eqref{eq:dm_vel} is equivalent to mapping the frequencies as $f\to f\del{1+\frac{v}{c}}$ in the coherent dedispersion kernel (Eq.~\ref{eq:kernel_cdd}).
Doppler-shifting the DM during coherent dedispersion is equivalent to applying the time-dependent geometric phase correction (which is equivalent to Lorentz transforming the data to the CHIME rest frame), $\boldsymbol{\phi}_{\text{geo}_0}$,  followed by coherent dedispersion to a reference DM (steps 2 and 3 Section \ref{sec:wideband_fringe_fitting}). We take the reference DM to be that measured at CHIME site, averaged over the \SI{\sim1.1}{\s} duration of the PSR B0531+21 sweep, and neglect the transformation to the SSB. This effect could be of an order of \SI{1e-4}{\parsec\per\centi\m\cubed} (above the required $\delta\text{DM}$ from fringe fitting), where other experiments such as the Pulsar Timing Array correct for it.

\subsection{Single baseline in VLBI and localization}
CHIME/FRB is capable of localizing FRBs up to \SI{\sim1}{\arcmin} with its own baseband localization pipeline \citep{2020arXiv201006748M}, and adding the extra VLBI baseline will provide constraints mostly in only one dimension (RA). Nevertheless, CHIME-ARO 10-m telescope localizations with $\sigma_\upalpha\sim\SI{200}{\mas}$ and $\sigma_\updelta\sim\SI{1}{\arcmin}$ may be sufficient to robustly associate an FRB with its host.
The upcoming CHIME/FRB Outriggers project will have three baselines that will further improve these localization errors.

The localization error will also affect the first guess of the phase geometric correction $\boldsymbol{\phi}_{\text{geo}_0}$ (and not more than a frame since its order will be less than \SI{2.56}{\micro\s} in our baseline), step 2  fringe fitting algorithm (Eqs. \ref{eq:baseband_geo_corr} and \ref{eq:phase_geo_correction}). Corrections in the case of an unknown target will be applied to the phase in order to find the true RA and DEC from the residual delay $\delta\tau$ (result from fringe fitting).

\subsection{CHIME/FRB Outriggers project}
The CHIME/FRB Outriggers project will have multiple CHIME single cylinders with two baselines thousands of kilometers from CHIME in order to observe and localize FRBs down to a \SI{50}{\mas} precision. Confirmed sites are: Allenby BC Canada, Green Bank Observatory WV USA, and Hat Creek Radio Observatory CA USA. For calibration the project will make use of pulsars (which have well known localizations of \SIrange{10}{20}{\mas}) as an alternative compared to steady-source calibrators, and because of CHIME's FoV the longest period of time without a pulsar on beam is roughly \SI{\sim1}{\hour}.
These outriggers will have tracking beams in each station, beamforming capabilities, and a triggering system, similar to the one implemented at the ARO 10-m telescope testbed. Correlation routines and methods described here will serve as a basis for the project. In particular, the outriggers beamforming capabilities will have pulsars available for calibration (most of the time), with a nearly in-beam calibration scheme \citep{2021AJ....161...81L}, by digitally pointing to target and references simultaneously (leaving only an sky angular distance interpolation), i.e., preserving coherence along the FRB dispersion time.

The project will also be interested in steady sources (VLBI calibrators) in order to account for the list of potential challenges described in this section. Such observations will enable us to characterize the VLBI network and measure our precision/uncertainties compared to the proposed pulsar calibration method. 

\section{Conclusions}
\label{sec:conc}
In the presented work we have developed a testbed for the CHIME/FRB Outriggers project with the capability of studying FRB localizations, using pulsars to coherently delay-reference, and demonstrated clock stability between an independent maser (ARO 10-m telescope) and GPS crystal oscillator (CHIME). The 1960s-era telescope at ARO was refurbished and updated to a modern system.
The 10-m dish is able to receive triggers from CHIME/FRB, record baseband data, and transfer it back to the cross-correlation site in a semi-autonomous process. We are currently recording single PSR B0531+21 pulses everyday (for potential calibration and clock stability) and in the future this will be increased to several pulses per day as well as other potential sources. The 10-m dish has also demonstrated reliability over extreme weather conditions, poor internet connections (satellite), and low maintenance on site during the 2020 COVID-19 pandemic.

Our worst localization scenario is $\theta\approx\SI{200}{\mas}$ (or equivalently \SIrange{-10}{10}{\nano\s}; baseline angle) in multiple-day observations, where only high signal-to-noise ratio bursts from the first five days in Figure \ref{fig:clock_jittering_md} were considered for this estimate. Other sources of localization error are: the on-sky separation between FRB (target) and calibrator (reference), the simultaneity of the observations, whether or not the ionospheric contribution is known in the FRB and calibrator directions, the clock jitter (CHIME and ARO 10-m telescope clock combination), and lastly whether or not assumptions in our correlator model remain true (Eq.~\ref{eq:vis_fit}).
Nevertheless, future calibrations (CHIME/FRB Outriggers project) will have access to pulsar (calibrators) $\lesssim\SI{1}{\hour}$ respect to the observed target (with the possibility of a nearly simultaneous beamformed baseband recording), plus comparing CHIME clock and DRAO maser can add an extra set of corrections to the clock (CHIME end; \citealt{2021arXiv211000576M}).
Clock corrections can be done by constantly monitoring calibrators in VLBI, in addition to using measurements of the DRAO maser to correct the clock jitter to the expected delay (given by calibrators). This leaves a comfortable window for the requirement of \SI{50}{\mas} proposed by the CHIME/FRB Outriggers project.

On the other hand, testbed delays over a single day behaved as expected given the precision of the CHIME and ARO 10-m telescope clock system, yielding the required precision of $\text{DM}\sim\SI{1e-8}{\parsec\per\centi\m\cubed}$ for a strong cross-correlation (below Doppler effect Section \ref{sec:geometric_delay_and_high_dm} and without any ionospheric prior in $\mathcal{L}$, Eq.~\ref{eq:likelihood-chi2}) and eventual localization.

Lastly, we showed proof of an FRB candidate cross-correlated with baseband data, providing enough signal-to-noise ratio and phase information to be considered as a true VLBI correlation. A proper and more in depth study of this event will be discussed in an upcoming paper.

\acknowledgments

We wish to acknowledge this land on which the University of Toronto operates. For thousands of years it has been the traditional land of the Huron-Wendat, the Seneca, and the Mississaugas of the Credit. Today, this meeting place is still the home to many Indigenous people from across Turtle Island and we are grateful to have the opportunity to work on this land.

We acknowledge that CHIME is located on the traditional, ancestral, and unceded territory of the Syilx/Okanagan people.

We wish to thank Rebecca Lin and Marten van Kerkwijk for the useful discussions on VLBI; and Gwendolyn Eadie, Joshua Speagle, and Luke Pratley for discussions on fringe fitting and statistical methods. We wish to thank all people involved in the telescope refurbishment and data acquisition: Dana Simard, James Willis, Vincent MacKay, Jos\'{e} Jauregui Garc\'{i}a, Jacob Taylor, and Nolan Denman. We wish to thank those useful discussions: Andrew Zwaniga, Emily Deibert and Victor Chan.

T.~Cassanelli is funded by the National Agency for Research and Development (ANID) / BECA DE DOCTORADO EN EL EXTRANJERO BECAS CHILE 2017 -- 72180183.

Computations were performed on the Niagara supercomputer at the SciNet HPC Consortium. SciNet is funded by: the Canada Foundation for Innovation; the Government of Ontario; Ontario Research Fund - Research Excellence; and the University of Toronto.

FRB research at MIT is supported by an NSF Grant (2008031).
FRB research at UBC is supported by an NSERC Discovery Grant and by the Canadian Institute for Advanced Research. 
We receive support from Ontario Research Fund-research Excellence Program (ORF-RE), Natural Sciences and Engineering Research Council of Canada (NSERC) [funding reference number RGPIN-2019-067, CRD 523638-201, 555585-20], Canadian Institute for Advanced Research (CIFAR), Canadian Foundation for Innovation (CFI), the National Science Foundation of China (Grants No. 11929301), Simons Foundation, THOTH Technology Inc., and Alexander von Humboldt Foundation. Computations were performed on the SOSCIP Consortium's [Blue Gene/Q, Cloud Data Analytics, Agile and/or Large Memory System] computing platform(s). SOSCIP is funded by the Federal Economic Development Agency of Southern Ontario, the Province of Ontario, IBM Canada Ltd., Ontario Centres of Excellence, Mitacs and 15 Ontario academic member institutions.

A.B.P.~is a McGill Space Institute (MSI) Fellow and a Fonds de Recherche du Qu\'ebec -- Nature et Technologies (FRQNT) postdoctoral fellow.
C.L.~was supported by the U.S. Department of Defense (DoD) through the National Defense Science \& Engineering Graduate Fellowship (NDSEG) Program.
D.M.~is a Banting Fellow.
D.C.G.~is supported by the John I.~Watters Research Fellowship
J.W.M.~is a CITA Postdoctoral Fellow: This work was supported by the Natural Sciences and Engineering Research Council of Canada (NSERC), [funding reference \#CITA 490888-16].
J.M.P.~is a Kavli Fellow.
K.B.~is supported by an NSF grant (2006548).
M.D.~is supported by a Killam Fellowship, Canada Research Chair, NSERC Discovery Grant, CIFAR, and by the FRQNT Centre de Recherche en Astrophysique du Qu\'ebec (CRAQ).
S.C.~acknowledges support from the NSF (AAG 1815242).
The CHIME/FRB baseband system is funded in part by a CFI John R.~Evans Leaders Fund award to I.H.S.
V.M.K.~holds the Lorne Trottier Chair in Astrophysics \& Cosmology and a Distinguished James McGill Professorship and receives support from an NSERC Discovery Grant and Herzberg Award, from an R.~Howard Webster Foundation Fellowship from the Canadian Institute for Advanced Research (CIFAR), and from the FRQNT Centre de Recherche en Astrophysique du Qu\'ebec.

\facilities{ARO 10-m telescope operated by the University of Toronto; and CHIME and the CHIME/FRB Collaboration at DRAO.}

\software{\texttt{astropy} \citep{2018AJ....156..123A}, \texttt{baseband} \citep{marten_van_kerkwijk_2020_4292543}, \texttt{difxcalc11} \citep{2016ivs..conf..187G}, \texttt{matplotlib} \citep{2007CSE.....9...90H}, \texttt{numpy} \citep{2020Natur.585..357H}, \texttt{scipy} \citep{2020SciPy-NMeth}, and \texttt{corner} \citep{Foreman-Mackey2016}.}


\appendix

\section{Radio Frequency chain}
\label{sec:rf_chain}

\begin{sidewaysfigure}
        \centering
        \includegraphics[scale=1]{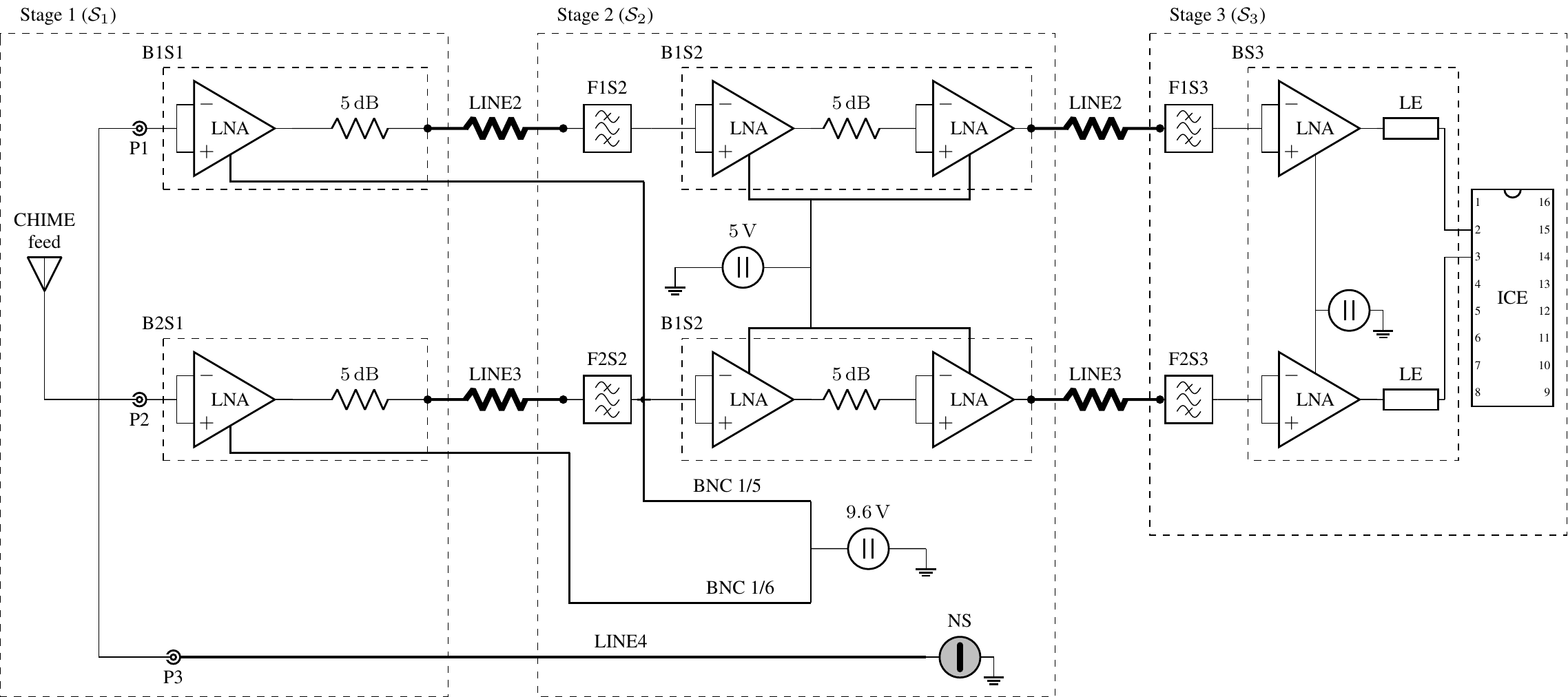}
        \caption{RF chain at the ARO 10-m telescope. Large dashed boxes are stages, and small dashed boxes are LNA assembled boxes. The amplification signal is a combination of CHIME custom and commercial amplifiers, and both polarizations undergo the same devices and cabling with the exception of LINE4. Thin, thick and very thick lines are SMA, BNC and N-type connections. LINE2 and LINE3 carry polarization information from the feed (left-hand side) to ICE (right-hand side), and LINE4 is a noise injection for the CHIME feed. Changes in the RF chain were done 2 years ago and it has been stable since then.}
        \label{fig:rf_chain}
\end{sidewaysfigure}

The RF chain at the ARO 10-m telescope has three main stages: telescope focus ($\mathcal{S}_1$), telescope pedestal ($\mathcal{S}_2$), and main control room ($\mathcal{S}_3$). Each stage is equipped with LNAs for the two polarizations. Long lines that go from the pedestal to the main control room are labeled LINE2 and LINE3 (N-type connection). Figure \ref{fig:rf_chain} contains all analog signal chain from the CHIME feed up to the digitalization at the ICE. Lines are extended for over \SI{200}{\m} from the pedestal to the main control room (right next to the 46-m dish), where the modernized equipment for the facility is located.
The lines and connection naming conventions have been kept from the original installation in the 1960s, and only SMA and N-type connections feed the RF chain. BNC provides power to the amplifiers. Underground lines LINE2 and LINE3 present a substantial loss and reflections (as seen in Figure \ref{fig:s_amps}), but with the current setup they fulfill the telescope purpose.

The amplification in the RF chain involves:
\begin{itemize}
    \item First stage amplifier boxes B1S1 and B2S1 are connected to the feed and contain a CHIME custom amplifier \citep{10.1117/12.2054950} and a \SI{5}{\decibel} attenuator to avoid reflections and feedback loops.
    \item Second stage amplifier boxes B1S2 and B2S2 have two commercial amplifiers, Mini-Circuits ZX60-112LN+ \citep{ZX60-112LN}, and a \SI{5}{\decibel} attenuation between them.
    \item Third stage amplifier box BS3 is located after \SI{200}{\m} of underground cabling, and the amplification is done by CHIME Pathfinder custom amplifiers \citep{10.1117/12.2054950}. The third stage ends with a line equalizer TLE LE7 400-800(29) S50.
\end{itemize}
Each bandpass filter in stages one and two is a Mini-Circuits ZABP-600-1+ \citep{ZABP-598-S}.
The system temperature of the telescope has not been properly computed so far. Future measurements of RF chain sections and sky observations will provide this information.

\section{Visibility and cross-correlation with baseband data}
\label{sec:visibility_and_cross-correlation_in_baseband_space}
The defined data format from Section \ref{sec:baseband_data} in the CHIME and ARO 10-m telescope baseband data, $\mathbf{V}\sbr{n, k}$, corresponds to a matrix of frequency channels (\SI{390.625}{\kilo\hertz} channel width), frames (\SI{2.56}{\micro\s} frame width), and two polarizations (either linear or circular basis).

\begin{figure*}[h]
    \centering
    \includegraphics{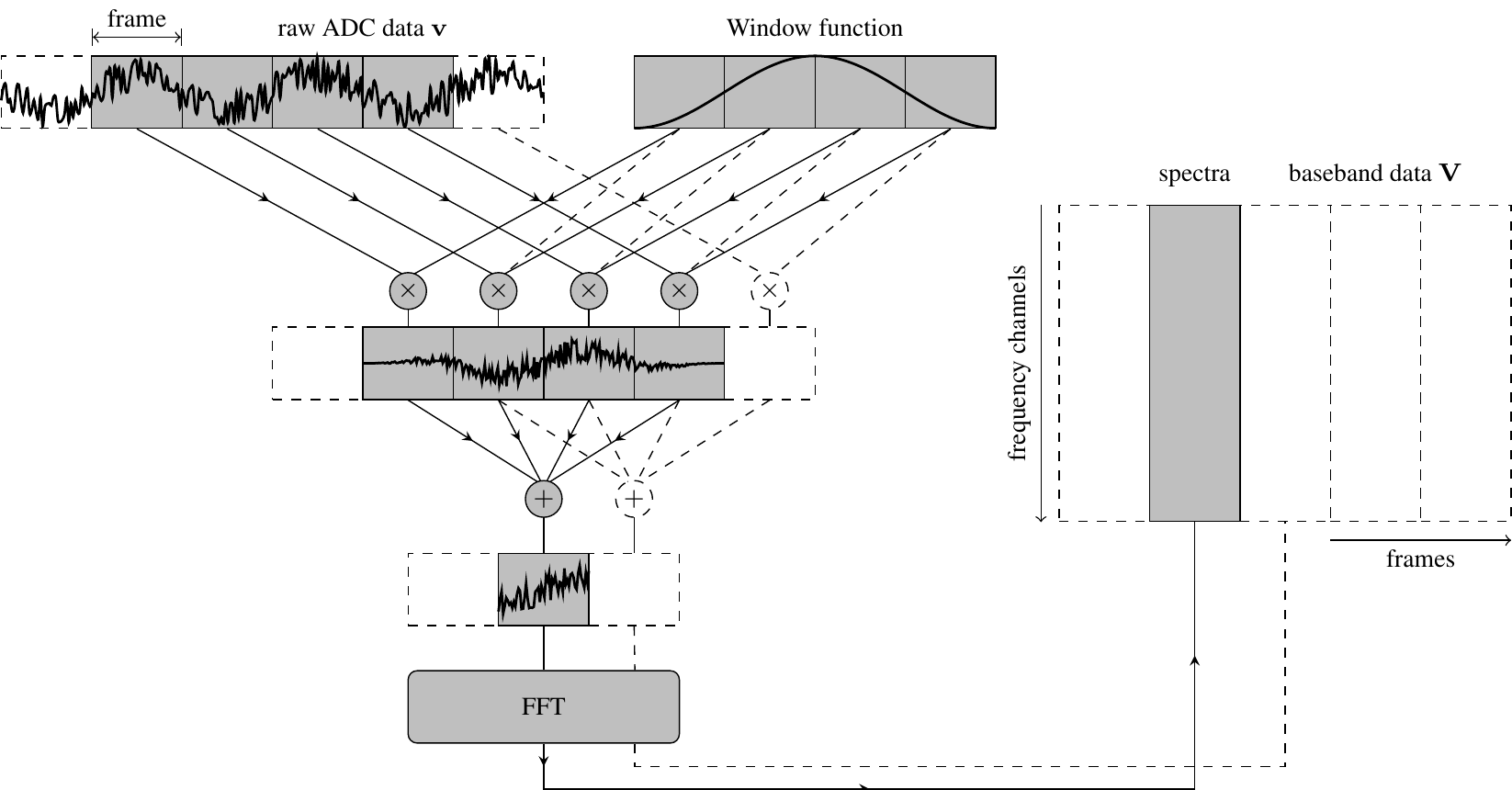}
    \caption{Graphical representation of the transformation of digitized raw ADC data $\mathbf{v}$ (with real-valued elements $v_q$) to baseband data $\mathbf{V}$ passing through a Polyphase Filter Bank. The discrete wavefront $\mathbf{v}$ (of $Q$ samples) is first divided in four sections of a frame size, then multiplied by a window function (of four taps) and summed over the frame size. The summed portion has the same number of points as a frame (\num{2048} points with \SI{1.25}{\nano\s} sample size). The summed section is then Fourier Transformed (FFT over time), where its spectra is then placed as a column of \num{1024} complex data points of the baseband data $\mathbf{V}$. The first iteration is shaded to gray, and the second iteration (dashed lines) will have another four (taps) frames (removing the initial left side and adding an extra side frame to the right).}
    \label{fig:pfb}
\end{figure*}

The process to form baseband data from a wavefront follows:
Once the wavefront ($v(t)$) is detected in the feed, the analog signal travels through until reaching the ICE system. Here it is digitized to discrete raw ADC data $\mathbf{v} = v_q \in\mathbb{R}^{Q}$ (with $Q$ an arbitrary total data points of samples but larger than \num{2048}) with a sample size of \SI{1.25}{\nano\s}, a purely real quantity. This digitized wavefront is then channelized (to form baseband data) through a Polyphase Filter Bank (PFB) \citep{2016arXiv160703579P, 2011PASA...28..317H} of four taps. Figure \ref{fig:pfb} shows a simplistic graphical representation of a PFB.
When the a frame size of \num{2048} samples is reached ($2048\times\SI{1.25}{\nano\s}=\SI{2.56}{\micro\s}$) the frame is multiplied by a section of a window function of the same number of points, the process repeats until all four taps are completed. Then the taps are averaged together yielding \num{2048} points which then pass through a one-dimensional FFT.
From the complex channelized spectrum half of it is discarded since the Fourier Transform of a real-valued function is Hermitian and placed as a column of $N=1024$ frequency channels in the baseband data $\mathbf{V}$.
The entire process is again repeated to form the following frame, where the first frame (first tap) used is discarded and a new frame is added.
In general, the PFB method is used to reduce sidelobes and minimize the leaked power between frequency channels.

Although the original wavefront changed its initial form, all physical information is still contained in the phase of the complex elements in the $\mathbf{V}$ matrix, viz., $\boldsymbol{\phi}=\text{Arg}\sbr{\mathbf{V}}$. Since a PFB has been applied (recorded data from ARO 10-m telescope and CHIME only return baseband data) going back to raw ADC data would imply undoing the transformation which is, in principle, revertible but not simple to accomplish. Nevertheless, working in baseband data does not affect the performance of the presented VLBI experiment.

Summary of data definitions follow:
\begin{itemize}
    \item \textbf{Raw ADC data} $\mathbf{v}_P = \del{v_P}_q\in\mathbb{R}^Q$: de-channelized real-valued raw voltage, before PFB, with element called sample \SI{1.25}{\nano\s}.
    \item \textbf{Baseband data} $\mathbf{V}_P=\del{V_P}_{nk}\in\mathbb{C}^{N\times K}$: channelized complex-valued raw voltage, after PFB, with element called frame \SI{2.56}{\micro\s}.
\end{itemize}
Here, $P$ represents the polarization.

\begin{figure*}
    \centering
    \includegraphics{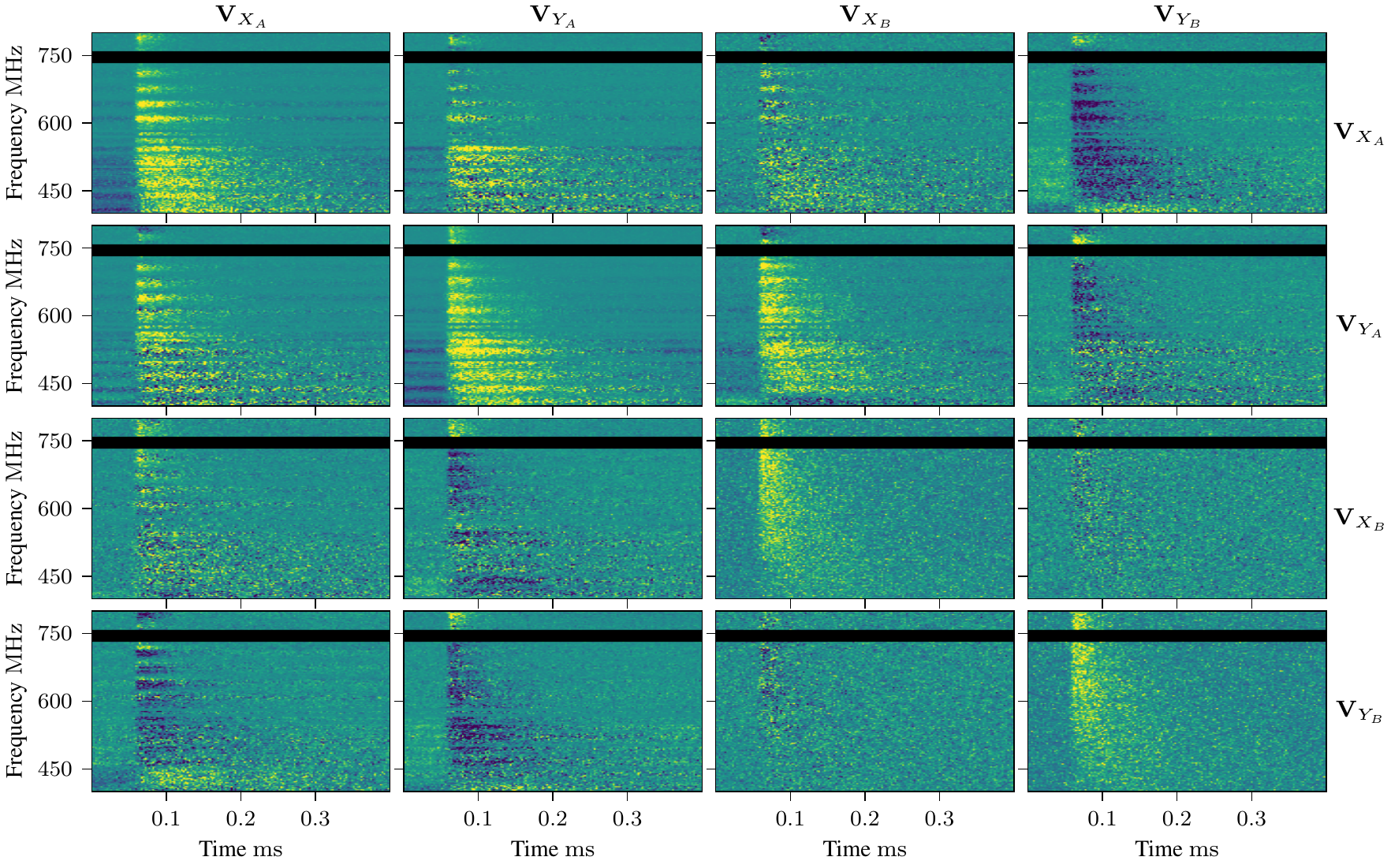}
    \caption{Visibility pairs for PSR B0531+21. The diagonal represents the intensity of PSR B0531+21 pulse for each polarization at each site $A$ and $B$. Off-diagonal section are real part (upper triangle) and imaginary part (lower triangle) of the computed visibilities. The pair of pulses corresponds to the same as in Figure \ref{fig:intensity_comparison} and referenced to the CHIME timestamp 2020-10-22 11:22:08.399687685. Visualization is done by taking $\mu\pm3\sigma$ from the amplitude value in each panel, and they are not normalized to each other.}
    \label{fig:correlation_matrix}
\end{figure*}

To find the cross-correlation function $\rho_{A, B}^\text{sf}(\tau)$ (Section \ref{sec:correlation_with_baseband_data}) in baseband data, we need to first form the visibility $\boldsymbol{\mathcal{V}}_{P_AP_B}$, which is nothing more than the element-by-element multiplication of two (same sized) matrices of baseband data: $\boldsymbol{\mathcal{V}}_{P_AP_B} = \mathbf{V}_{P_A}\overline{\mathbf{V}_{P_B}}$ (same as in Eq.~\eqref{eq:visibility}), with $P_A$ and $P_B$ the polarization basis of each site.
The correct ``frame alignment'' between the two baseband pairs $\del{\mathbf{V}_{P_A}, \mathbf{V}_{P_B}}$ must be perfectly aligned within a frame (as in frame from site $A$ timestamp corresponds to the frame from site $B$ timestamp within \SI{2.56}{\micro\s}, applying the necessary corrections $k_\text{shift}$ Eq.~\eqref{eq:k_shift}, Section \ref{sec:wideband_fringe_fitting} forming visibilities step 5), otherwise the visibility will not correlate and signal will vanish (even if only one frame off).
From a single pair of pulses (e.g., those in Figure \ref{fig:intensity_comparison}), we can generate four visibilities given by the polarization pairs: $(\mathbf{V}_{X_A}, \mathbf{V}_{X_B})$; $(\mathbf{V}_{X_A}, \mathbf{V}_{Y_B})$; $(\mathbf{V}_{Y_A}, \mathbf{V}_{X_B})$; and $(\mathbf{V}_{Y_A}, \mathbf{V}_{Y_B})$. Figure \ref{fig:correlation_matrix} shows a matrix of all possible combinations of visibilities, in cross-correlation and within single sites (i.e., $X$ and $Y$ pairs from the same site). The matrix diagonal is the pulse autocorrelation of single polarizations $\mathbf{V}_{P_k}\overline{\mathbf{V}_{P_k}}$ with $k = A, B$ and $P=X, Y$. The off-diagonal panels are the real part (upper section) and imaginary part (lower section), $\text{Re}\sbr{\boldsymbol{\mathcal{V}}_{P_AP_B}}$ and $\text{Im}\sbr{\boldsymbol{\mathcal{V}}_{P_AP_B}}$, respectively. The set of shown visibilities went through a partial fringe fitting (Section \ref{sec:wideband_fringe_fitting}) until step 5.

All analyses done over single and multiple days, Sections \ref{sec:clock_stability_single_day} and \ref{sec:clock_stability_multiple_day}, were performed with a single pair of polarizations $\mathbf{V}_{Y_A}$ and $\mathbf{V}_{X_B}$, i.e., the one with the strongest cross-correlation and most stable phase.
The imaginary part of the visibility $\boldsymbol{\mathcal{V}}_{Y_AX_B}=\mathbf{V}_{Y_A}\overline{\mathbf{V}_{X_B}}$ is plotted in Figure \ref{fig:phase_baseband}.

\bibliography{frb-vlbi-10m-ref}
\bibliographystyle{aasjournal}

\end{document}